\documentclass[11pt]{article}
\usepackage{times,fancyhdr}
\usepackage{cite}
\usepackage{graphicx}

\title{Toward an adequate mathematical model of mental space:
conscious/unconscious dynamics on $m$-adic trees}

\author{Andrei Yu. Khrennikov\\
Center for Mathematical
Modeling \\ in Physics and Cognitive Sciences,\\
University of V\"axj\"o, S-35195, Sweden\\
Email:Andrei.Khrennikov@msi.vxu.se}

\begin{document}

\maketitle

\abstract{We try to perform geometrization of cognitive science and
psychology by representing information states of cognitive systems
by points of {\it mental space} given by a hierarchic $m$-adic tree.
Associations are represented by balls and ideas by collections of
balls. We consider dynamics of ideas based on lifting of dynamics of
mental points. We apply our dynamical model for modeling of flows of
unconscious and conscious information in the human brain. In a series
of models, Models 1-3, we consider cognitive systems with increasing
complexity of psychological behavior determined by structure of
flows of associations and ideas.}

\medskip

Keywords: Mental space, hierarchic encoding of mental information,
m-adic trees and numbers,  dynamical systems, conscious/unconscious
flows of information. neuronal trees, psychoanalysis

\section{Introduction}

One of the sources of the extremely successful mathematical
formalization of physics was creation of the adequate mathematical
model of physical space, namely, the Cartesian product of real
lines. This provides the possibility for ``embedding''  physical
objects into a mathematical space. Coordinates of physical systems
are given by points of this space. Rigid physical bodies are
represented by geometric figures (cubes, balls,... ). By describing
dynamics of coordinates, e.g., with the aid of differential
equations, we can describe dynamics of  bodies (from falling stones
to Sputniks).

In a series of works (Khrennikov, 1997, 1998a,b, 1999a,b, 2000a,b)
there was advocated a similar approach to description of mental
processes in cognitive sciences and psychology (and even information
dynamics in genetics), see also Albeverio et al., 1999, and
Dubischar et al, 1999. Similar to physics, the first step should
be elaboration of a mathematical model of {\it mental space.} We
understood well that this is a problem of huge complexity and it
might take a few hundred years for creating an adequate
mathematical model of mental space. We recall that it took three
hundred years to create a mathematically rigorous model of real
physical space. In previous papers critical
arguments were presented  against the real model of space as a possible candidate
for a  mental space. One of the main arguments was that the real
continuum is a continuous infinitely divisible space.  Such a
picture of space is adequate to physical space (at least in
classical physics), but {\it mental space is not continuous:} mind
is not infinitely divisible! Another problem with the real continuum
is that it is homogeneous: ``all points of this space have equal
rights.'' In opposition to such a homogeneity mental states have
clearly expressed {\it hierarchical structure,} see for discussions:
Hubel and Wiesel, 1962, Smythies, 1970, Clark, 1980, Amit, 1989, Bechtel and Abrahamsen, 1991,
Khrennikov, 1997, 1998a,b, 1999a,b, 2000a, b; Albeverio et
al., 1999; Dubischar et al, 1999, Voronkov, 2002a, b, Sergin,  2007,  see also
Bechterew, 1911, Damasio et al., 1989, Fuster,  1997, for
corresponding medical evidence.

Therefore a model of mental space that we are looking for should be
(at least)  {\it discontinuous and hierarchical.} In mathematics
there is a well known class of spaces with such features. These are
$m$-adic trees (here $m$ is a natural number giving the number of
branches of a tree at each vertex). It is interesting that such
trees are nicely equipped: there is a well defined algebraic
structure which gives the possibility to add, subtract, multiply,
and for prime $m$ (so $m=2,3,5,..., 1997, 1999,...)$ even divide
branches of such a tree. There is a natural topology on such trees
encoding the hierarchic tree  structure. This topology is based on a
metric, so called {\it ultrametric.} Thus $m$-adic trees are not
worse equipped than the real line. However, the equipment -- algebra
and topology -- is very different from the real one.

We proposed (see Khrennikov, 1997, 1998a,b, 1999a,b, 2000a,b;
Albeverio et al., 1999; Dubischar et al, 1999) to choose $m$-adic
trees as possible models of {\it mental space $X_{\rm{mental}}$:}
 points of this space are branches of a tree.
These are {\it mental coordinates} representing {\it mental states.}
By using mental coordinates we able to embed into the space mental
analogs of physical rigid bodies -- {\it associations and ideas.}
They are represented, respectively, by {\it balls and collections of
balls} in the ultrametric mental space.

Mental states (represented by branches) are {\it basic cognitive mental images.} An
association connects a number of cognitive mental images. Thus an
association can be represented as a subset of the mental space. The
crucial point is that in our model the associative connection of
cognitive mental images is fundamentally hierarchical. Therefore an
association is not an arbitrary collection of cognitive mental images
(not an arbitrary set of mental points), but a hierarchically
coupled collection. Since in our model the mental hierarchy is encoded by the
topology of the mental space, it represents the associative coupling
of cognitive mental images into balls. A larger ball couples together
more cognitive mental images. Thus it is a more complex association
(but it is a ``fuzzy-association,'' it is not sharp).
Decreasing of ball's radius induces decreasing of the complexity an
association which is represented by this ball. An association becomes sharper.
In the limit we
obtain the ball of the zero radius. That is nothing else than a
single mental point (the center of such a degenerated ball). 
This is a single cognitive mental image. This is the limiting case
of an association: a cognitive mental image is "associated" with
itself. We hope that such a limiting degeneration of an association
into a mental image would not be misleading for readers.

Ideas are identified as collections of associations, something
analogous to a coherent group of individuals in the biological
analogy. The identification of the  fundamental  structure as a
mental image allows a concrete dynamical model for ideas as
collections of loosely bound associations. Association is kind of
atom of cognition from which more complex ideas are build like
molecules from atoms.

We mention also that a $p$-adic model (here $m=p$ is a prime number)
of consciousness was (independently) proposed in Pitk\"anen, 1998.
Pitk\"anen's approach was not based on encoding of {\it mental
hierarchy} by $p$-adic numbers. It has a deeper relation to
foundations of physics, especially the quantum one.

Recently $p$-adic information space was used for genetic models, see
Dragovich et al., 2006, and Khrennikov, 2006a,b as well as 
Pitk\"anen,  2006. A new exciting domain of research is use
of ultrametric methods in data-analysis -- from 
astrophysics and computer science to biology, see, e.g.,  Murtagh, 2004.

In papers, Khrennikov, 1997, 1998a,b, 1999a,b; Albeverio et al.,
1999; Dubischar et al, 1999,  we studied merely the dynamics of
mental states -- mental images.  We considered dynamical systems
which work with mental states.  There is a nonlinear relation
between input and output mental states,
\begin{equation}
\label{e1} x_{n+1}=f(x_n), x_n \in X_{\rm{mental}}.
\end{equation}
The description of functioning of the human brain by dynamical
systems (feedback processes) is a well established approach. The
main difference between our approach and the {\it conventional
dynamical approach to cognition} (see Ashby, 1952, van Gelder and
Port, 1995, van Gelder, 1995, Strogatz, 1994, Eliasmith, 1996, Conte
et al, 2006 -- in the latter there was presented a dynamical model of
cognition exhibiting nondeterministic features similar to those in
quantum mechanics) is that in the conventional
dynamical approach  dynamical systems work in
the real physical space of electric potentials and in our approach
dynamical systems work in the $m$-adic mental space. There is also a
similarity between our approach and the artificial intelligence
approach, Chomsky, 1963, Churchland and  Sejnovski, 1992.

In the present paper we study dynamics of mental analogs of
physical rigid bodies -- associations (balls in the mental metric
space) and ideas (collections of balls). In spite of the fact that
dynamics of associations and ideas can be in principle reduced to
dynamics of mental points composing those ``mental bodies'', the
those dynamics exhibits their own interesting properties which could
not be seen on the level of pointwise dynamics.

We apply our dynamical model for modeling of flows of unconscious
and conscious information in the human brain.\footnote{We do not try
to discuss general philosophical and cognitive problems of modeling
of consciousness, see, e.g., Blomberg et al. 1994, Baars, 1997,
Pitk\"anen, 1998, Khrennikov, 1998a, 2004a.}

In series of models, Models 1-3, we consider cognitive systems with
increasing complexity of psychological behavior determined by
structure of flows of associations and ideas. Using this basic
conceptual repertoire an increasingly refined cognitive model is
developed starting from an animal like individual whose sexual
behavior is based on instincts alone. At the first step a
classification of ideas to interesting and less interesting ones is
introduced and less interesting ideas are deleted. At the next level
a censorship of dangerous ideas is introduced and the conflict
between interesting and dangerous leads to neurotic behaviors, fix
idees, and hysteria. As was pointed out by one of referees of this
paper, these aspects of the model reflect more the general structure
of conscious/unconscious processing rather than properties of
$m$-adic numbers. The basic mathematical structure for this model is
mental ultrametric space. In particular, ultrametric is used to
classify ideas -- to assign to each idea its measures of interest and interdiction.

Finally, we apply our approach to mathematical modeling of Freud's
theory, see, for example, Freud, 1933, of interaction between
unconscious and conscious domains. One of basic features of our
model is splitting  the process of thinking into two separate (but
at the same time closely connected) domains: {\it conscious} and
{\it unconscious}, cf. Freud, 1933.  We shall use the following
point of view on the simultaneous work of the consciousness and
unconsciousness. The consciousness contains a {\it control center}
$CC$ that has functions of control over results of functioning of
subconsciousness. $CC$ formulates problems, and sends them to the
unconscious domain. The process of finding a solution is hidden in
the unconscious domain. In the unconscious domain there work
gigantic dynamical systems -- {\it thinking processors.} Each
processor is determined by a function $f$ from mental space into
itself (describing the corresponding feedback process --
psychological function). It produces iterations of mental states
(points of mental space) $x_1=f(x_0),...., x_{n+1}= f(x_n),...$
These intermediate mental states are not used by the consciousness.
The consciousness (namely $CC)$ controls only some exceptional
moments in the work of the dynamical system in the unconscious
domain -- attractors and cycles. Dynamics of mental points induce
dynamics of mental figures, in particular, ball-associations and,
hence, ideas (collections of balls). The crucial point is that
behaviors of the dynamical in the mental space and its lifting to
spaces of associations and ideas can be very different. Extremely
cycling (chaotic) behavior on the level of mental states can imply
nice stabilization to attractors on the level of ideas.

To couple our hierarchic  $m$-adic model of processing of
information in brain with other investigations in cognitive brain
research, we can mention  e.g. the paper of Oztop et al, 2005,
presenting an approach to ``mental state inference" (oriented toward
visual feedback control). We remark that "mental state inference",
more generally ``theory of mind" has been a resent topic of interest
in cognitive neuroscience, see Blakemore and Decety, 2001, Chaminade
et al., 2001,  Frith C.D. and Frith U., 1999. In fact, we present a
model of mental state inference serving for Freud's psychoanalysis.
Our model is about estimation of mental states on the basis of
symptoms (in the mentioned papers mental states were estimated on
the behavioral basis).

Geometrically we can imagine a system of $m$-adic integers  (which
will be the mathematical basis of our cognitive models) as a
homogeneous {\it tree} with $m$-branches splitting at each vertex.
 The distance between mental states is determined by
the length of their common root: close mental states have a long
common root.  The corresponding geometry strongly differs from the
ordinary Euclidean geometry.
\begin{center}
\begin{figure}[ht] \unitlength1cm
\begin{picture}(12,5)
\put(2,3){\circle{0.5}}            \put(1.9,2.9){$\star$}
\put(2.5,2.8){\vector(2,-1){2.0}} \put(2.5,3.2){\vector(2,1){2.0}}
\put(5,4.4){\circle{0.5}}            \put(4.9,4.28){$0$}
\put(5.5,4.35){\vector(4,-1){2.0}} \put(5.5,4.45){\vector(4,1){2.0}}
\put(5,1.6){\circle{0.5}}            \put(4.9,1.48){$1$}
\put(5.5,1.55){\vector(4,-1){2.0}} \put(5.5,1.65){\vector(4,1){2.0}}
\put(8,5.05){\circle{0.5}}          \put(7.9,4.95){$0$}
\put(8,3.7){\circle{0.5}}          \put(7.9,3.57){$1$}
\put(8,2.25){\circle{0.5}}          \put(7.9,2.15){$0$}
\put(8,0.9){\circle{0.5}}          \put(7.9,0.75){$1$}
\put(8.5,5.00){\vector(4,-1){1.0}} \put(8.5,5.10){\vector(4,1){1.0}}
\put(8.5,3.65){\vector(4,-1){1.0}} \put(8.5,3.75){\vector(4,1){1.0}}
\put(8.5,2.20){\vector(4,-1){1.0}} \put(8.5,2.30){\vector(4,1){1.0}}
\put(8.5,0.85){\vector(4,-1){1.0}} \put(8.5,0.95){\vector(4,1){1.0}}
\end{picture}
\caption{The $2$-adic tree}
\end{figure}
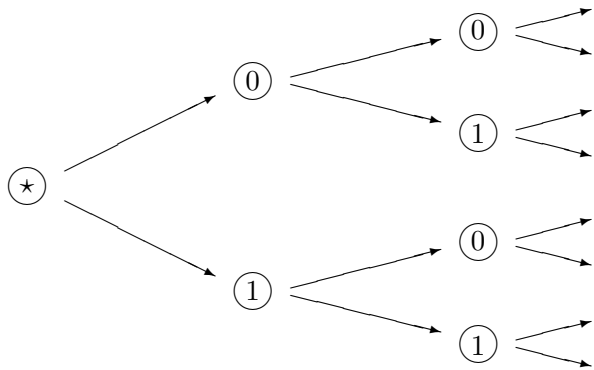
\end{center}

We also point out that systems
of $m$-adic numbers (restricted to $m=p$ a prime number) were
intensively used in theoretical physics, see e.g. Khrennikov, 1997.

We remark that in this paper we do not consider in details the neuronal basis
of the $m$-adic mental space, see Khrennikov 2004a, b, for
corresponding models, see section 5 for  a brief review.
  This neuronal basis is provided by
consideration of hierarchical neuronal trees. Such trees provides
connection of the mental space (produced by a tree) with physical
space (in that neuronal trees are located). Mental processes are
connected with physical and chemical processes in the brain: mental
states are produced as distributed activations of neuronal pathways.
We remark that mental hierarchy  was discussed  a lot, see already
mentioned papers Smythies, 1970, Clark, 1980, Amit, 1989,  Hubel and Wiesel, 1962, 
Bechtel and Abrahamsen, 1991, Ivanitsky, 1999, Watt  and Phillips, 2000,
Stringer and Rolls,  2002, Khrennikov, 1997, 1998a,b, 1999a,b, 2000a, b; Albeverio et
al., 1999; Dubischar et al, 1999, Voronkov, 2002a, b, Sergin,  2007.
However, there were not so much experimental neurophysiological evidences
of existence of such neuronal hierarchical structures in the brain.
Therefore the recent paper of Luczak et al., 2007, that 
confirmed experimentally existence of neurons-directors which rule
the performances of cognitive tasks (under the same context and
learning conditions) is extremely important for our model.
The presence of a complicated hierarchy of time scales in the brain can be considered 
as an indirect confirmation of the hierarchical structure of processing of information 
in the brain, see, e.g., Geissler et al (1978),
Geissler and Puffe (1982), Geissler (1983, 85, 87,92), Geissler and
Kompass (1999, 2001), Geissler, Schebera, and Kompass (1999), Klix and van der Meer (1978),
Kristofferson (1972, 80, 90),  Bredenkamp (1993), Teghtsoonian
(1971).

We start with recollection of the basic notions from e.g.  
Khrennikov, 1997, 1998a,b, 1999a,b, 2000a,b; Albeverio et al., 1999;
Dubischar et al, 1999.  Consequences for neurophysiology,  neuroinformatics,
and cognitive sciences as well as for psychology and neuropsychology, and even 
medicine and psychiatry will be presented in sections 11--13; possibilities to apply our 
model for the project on artificial life will be considered in section 14 
(``psychological robots'').

\section{$m$-adic ultrametric spaces}

The notion of a metric space is used in many applications for
describing distances between objects. Let $X$ be a set. A function
$\rho: X\times X \to {\bf R}_+ $ (where ${\bf R}_+ $ is the set of
positive real numbers) is said to be a {\it metric} if it has the
following properties: $ 1) \rho(x,y)= 0 \; \; \mbox{iff}\;  x=y \; \;
\mbox{(non-degenerated)};  2)\rho(x,y)=\rho(y,x)
\;\;\mbox{(symmetric)}; 3) \rho(x,y)\leq \rho(x,z) +\rho(z,y)\; \;
\mbox{(the triangle inequality).}$  The pair $(X,\rho)$ is called a
metric space.

We are interested in the following class of metric spaces
$(X,\rho).$ Every point $x$ has the infinite number of coordinates
\begin{equation}
\label{K} x=(\alpha_1,...,\alpha_n,...)\;.
\end{equation}
Each coordinate yields the finite number of values,
\begin{equation}
\label{K1} \alpha\in A_m=\{ 0,...,m-1\},
\end{equation}
where $m > 1$ is a natural number, the base of the alphabet $A_m.$
The metric $\rho$ should be so called ultrametric, i.e.,
satisfy the {\it strong triangle inequality}:
\begin{equation}
\label{K2} \rho (x,y)\le\max [\rho (x,z),\rho (z,y)], \; x,y,z\in X.
\end{equation}
The strong triangle inequality can be stated geometrically: {\it
each side of a triangle is at most as long as the longest one of the
two other sides.} Such  a triangle is quite restricted when considered in
the ordinary Euclidean space.

We denote the space of sequences (\ref{K}), (\ref{K1}) by the symbol
${\bf Z}_m.$ The standard ultrametric is introduced on this set in
the following way. For two points

$x=(\alpha_0, \alpha_1, \alpha_2,...., \alpha_n,.....),
y=(\beta_0,\beta_1, \beta_2, ..., \beta_n,...)\in {\bf Z}_m,$

we set
$$
\rho_m(x,y)= \frac{1}{m^k} \; \; \mbox{if}\; \;  \alpha_j= \beta_j,
j=0,1,...,k-1,\; \;  \mbox{and} \; \;\alpha_k\not=\beta_k.
$$
This is a metric and even an ultrametric.  To find the distance
$\rho_m(x,y)$ between two strings of digits $x$ and $y$ we have to
find the first position $k$ such that strings have different digits
at this position.

Let $(X,\rho)$ be an arbitrary ultrametric space. For  $r\in{\bf
R}_{+},a\in X,$ we set
  $$U_r(a)=\{ x\in X:\rho(x,a)\le r \},\; \;
  U_r^-(a)=\{ x\in X:\rho(x,a) < r\} .
  $$
  These are {\it balls}
  of radius $r$ with center $a.$  Balls have the following
  properties,  Khrennikov, 1997:

1) Let $U$ and $V$ be two balls in $X.$ Then there are only two
possibilities: (a) balls are ordered by inclusion (i.e., $U\subset
V$ or $V\subset U$); (b) balls are disjoint.\footnote{There is the
third possibility in the Euclidean space .}

2) Each point of a ball may serve as a centre.

3) In some ultrametric spaces a ball may have infinitely many radii.

Let $m > 1$ be the fixed natural number. We consider the $m$-adic
metric space $({\bf Z}_m, \rho_m).$  This metric space has the
natural algebraic structure, see  Khrennikov, 1997.

A point $x=(\alpha_0, \alpha_1, \alpha_2,..., \alpha_n,....)$ of the
space ${\bf Z}_m$ can be identified with a so called $m$-adic number:
\begin{equation}
\label{l1} x= \alpha_0 \alpha_1...\alpha_k....\equiv \alpha_0
+\alpha_1 m +...+ \alpha_k m^k +... \;.
\end{equation}
The series (\ref{l1}) converges in the metric space ${\bf Z}_m.$ In
particular, a finite mental string $x= \alpha_0 \alpha_1...\alpha_k$
can be identified with the natural number
$$
x= \alpha_0+  \alpha_1 m+...+\alpha_k m^k.
$$
Therefore the set of all finite mental strings can be identified
with the set of natural numbers ${\bf N}.$ So dynamics of finite
mental strings can be simulated via dynamics on ${\bf N}.$ Moreover,
${\bf N}$ is a dense subset of ${\bf Z}_m:$ each $x \in  {\bf
Z}_m$ can be approximated with an arbitrary precision by natural
numbers. Thus the space $m$-adic numbers can be considered as an
extension of the set of natural numbers. By choosing different $m$
we obtain in general different extensions ${\bf Z}_m.$ Therefore we
can say that {\it in our model mental states are encoded by natural
numbers.} However, the mental geometry on the set of natural numbers
differs crucially from the one which is induced from the real line.

It is possible to introduce algebraic operations on the set of
$m$-adic numbers ${\bf Z}_m,$ namely addition, subtraction, and
multiplication. These operations are natural extensions by the
$m$-adic continuity of the standard operations on the set of natural
numbers ${\bf N}=\{ 0,1,2,3,...\}.$

\section{ Mental Space}

We shall use the following mathematical model for mental space:

\medskip

(1) Set-structure: The set of mental states $X_{\rm{mental}}$ has
the structure of the $m$-adic tree: $X_{\rm{mental}}= {\bf Z}_m.$

\medskip

(2) Topology: Two mental states $x$ and $y$ are close if they have
sufficiently long common root. This topology is described by the
metric $\rho_m.$

\medskip

In our mathematical model {\it mental space is represented as the
metric space $({\bf Z}_m, \rho_m).$}

Dynamical thinking  on the level of mental states is performed via
the following procedure: an initial mental state $x_0$ is sent to
the unconscious domain; it is iterated by some dynamical system
which is determined by a map
$$
f: {\bf Z}_m \to {\bf Z}_m;
$$
an attractor is communicated to the consciousness; this is the
solution of a problem $x_0.$\footnote{ It may be that iterations
starting with some $x_0$ will not arrive to any attractor. For
example, starting with $x_0,$ $\tau$ may perform a cyclic  behavior
in the process of thinking. In such case a cognitive system $\tau$
would not find the definite solution of a problem. In particular, it
is impossible to escape a cyclic behavior on the level of mental
states: even the simplest dynamical systems in ${\bf Z}_m$ may have a
huge number of cycles, see Khrennikov, 1997.} Our mathematical model
is based on two cornerstones:

\medskip

H).  The first is the assumption that the coding system which is
used by the brain for recording vectors of information generates a
{\it hierarchical structure} between digits of these vectors. Thus
if $x=(\alpha_1, \alpha_2,...,\alpha_n,...), \alpha_j=0,1,...,m-1,$
is an information vector which presents in the brain a mental state
then digits $\alpha_j$ have different weights. The digit $\alpha_0$
is the most important, $\alpha_1$ dominates over $\alpha_2,...,
\alpha_n,...,$ and so on.

D). The second is the assumption that functioning of the brain is
not based on the {\it rule of reason}. The unconsciousness is a
collection of dynamical systems $f_s(x)$ (thinking processors) which
produce new mental states practically ~automatically. The
consciousness only uses and control results (attractors in spaces of
ideas) of functioning of unconscious processors.

For a neuronal basis of the $m$-adic mental space, see Khrennikov
2004a, b. We also mention the possibility to apply the hierarchic
mental space to genetics.

{\bf Example 3.1} (4-adic genetic information space) We may describe
DNA and RNA sequences by m-adic numbers. We present schematically
development of this model. DNA and RNA sequences are represented by
4-adic numbers. Nucleotides are mapped to digits in registers of
4-adic numbers: adenine - $A,$ guanine - $G,$ cytosine - $C,$ and
thymine - $T$ are encoded by $\alpha=0,1,2,3.$ The $U$-nucleotide is
represented by 3. The {\it DNA and RNA sequences have the natural
hierarchic structure: letters which are located at the beginning of
a chain are considered as more important.} This hierarchic structure
coincides with the hierarchic structure of the 4-adic tree. It can
also be encoded by the 4-adic metric. The process of
DNA-reproduction is described by action of 4-adic dynamical system.
As we know, the genes contain information for production of
proteins. The genetic code is a degenerate map of codons to
proteins. We model this map as functioning of a monomial 4-adic
dynamical system. Proteins are attractors of this dynamical system.
We also can study the process the genom evolution in the framework
of 4-adic dynamical systems.

\section{Associations and
Ideas}

We now improve this dynamical cognitive model on hierarchic mental
trees by introducing a new hierarchy: {\it equivalence classes of
mental states are interpreted as associations, collections of
associations as ideas.}

A new property of dynamics of ideas is that (for a large class of
dynamical systems on $m$-adic trees) for each initial idea $J_0$ its
iterations  are {\bf attracted} by some idea $J_{\rm{attr}},$ see
Khrennikov, 1997, 2004a, for mathematical details. The latter idea
is considered by the consciousness as a solution of the problem
$J_0$. In the opposition to such an attractor-like dynamics of
ideas, dynamics of mental states (on the tree $X_{\rm{mental}}$) or
associations need not be attractive. In particular, there can exist
numerous cycles or ergodic evolution.

By using higher cognitive levels (associations and ideas) of the
representation of information a cognitive system strongly improves
the regularity of thinking dynamics. Finally, we note that the use
of a new cognitive hierarchy (in combination with the basic
hierarchy of the $m$-adic tree) strongly improves the information
power of a cognitive system.

Special collections of mental points form new cognitive objects,
{\it associations.} Let $s \in \{ 0,1,..., m-1\}.$ A set
$$
A_s=\{x= (\alpha_0,..., \alpha_k,...) \in {\bf Z}_m: \alpha_0=s\}
$$
is called an association of order 1. By realizing ${\bf Z}_m$ as
the metric space we see that $A_s$ can be represented as the ball of
radius $r=\frac{1}{m}.$  Any point $a$ having $\alpha_0$ as the
first digit can be chosen as a center of this ball (we recall that
in an ultrametric space any point $a$ belonging to a ball can be
chosen as its center): $A_s=U_{\frac{1}{m}}(a),$ where
$a=(a_0,a_1,...), a_0=s.$ Associations of higher orders are defined
in the same way. Let $s_0,..., s_{k-1} \in \{ 0,1,..., m-1\}.$ The
set
$$
A_{s_0...s_k} = \{x= (\alpha_0,..., \alpha_k,..) \in {\bf Z}_m:
\alpha_0=s_0,..., \alpha_{k-1}=s_{k-1}\}
$$
is called an association of order $k.$ These are balls of the
radius $r=\frac{1}{m^k}.$

Denote the set of all associations by the symbol $X_A.$ Collections
of associations will be called {\it ideas.} Denote the set of all
ideas by the symbol $X_{ID}.$ The space $X_{ID}$ consists of
points-associations.

In this section  we study the simplest dynamics  of associations and
ideas which are induced by corresponding dynamics of mental states,
\begin{equation}
  \label{i2}
  x_{n + 1} = f(x_n).
\end{equation}
Suppose that, for each association, its image  is again an
association. Thus $f$ maps balls onto balls.  Then dynamics
(\ref{i2}) of mental states of $\tau$ induces dynamics of
associations
\begin{equation}
  \label{i3}
  A_{n + 1} = f(A_n).
\end{equation}
We say that dynamics in the mental space $X_{\rm{mental}}$ for
transformations is lifted to the space of associations $X_A.$

\section{Neuronal Realization}

Let us consider the simplest model of a neuronal tree
$T_{\rm{neuronal}}$ inducing a mental space $X_{\rm{mental}}.$ This
model is based on the 2-adic neuronal tree given by Figure 1. Each
vertex of this tree corresponds to a single neuron. In this
idealized model each axon provides connections with precisely two
neurons of lower level of the hierarchy in the neuronal tree. There
is the root-neuron denoted by $\star,$ its axon provides connections
with the two neurons, 0 and 1, of the lower level. Each of these
neurons sends its axon to precisely two neurons of the lower level
and so on. Each branch $n$ of this tree ( a hierarchical chain of neurons) 
can be coded by a sequence of
zeros/ones  Thus this neuronal tree can be mathematically
represented as the set of 2-adic numbers: $T_{\rm{neuronal}}={\bf Z}_2.$

Each branch  of this  neuronal  tree is a device for producing
mental states (cognitive mental images). In the simplest model we
suppose that each neuron can be only in the two states: $\alpha=1,$
firing, $\alpha=0,$ non-firing. Thus each branch produces (at some
moment of time) a sequence of zeros/ones: $x=\alpha_0
\alpha_1...\alpha_N...,$ where $\alpha_j=0,1.$ (In the mathematical
model we can consider infinitely long sequences). Thus the neuronal
tree $T_{\rm{neuronal}}={\bf Z}_2$ produces the mental space
$X_{\rm{mental}}={\bf Z}_2.$ We can consider a {\it mental field} on
the neuronal tree $T_{\rm{neuronal}}.$ This is the map
$$
\psi: T_{\rm{neuronal}} \to X_{\rm{mental}},
$$
mathematically:
$$
\psi: {\bf Z}_2 \to {\bf Z}_2, \; \psi(n)=x.
$$

In fact, we need not assume that the 2-adic mental space should be
based on the 2-adic morphology of the neuronal tree. In general
there is no direct connection between the morphology of a the
neuronal tree and the corresponding mental space. Let us consider
any tree $T_{\rm{neuronal}}$  with the root  $\star$ (any number of
edges leaving a vertex, so an axon can provide connections with any
number of neurons at the lower floor). Nevertheless, let us consider
the same firing/not coding system. Each branch of the neuronal tree
$T_{\rm{neuronal}}$ produces a 2-adic number. Here the mental field
is a ${\bf Z}_2$-valued function on $T_{\rm{neuronal}}.$ This is an
important property of the model: it would be not so natural to
consider only homogeneous neuronal trees of the $m$-adic type.

The structure of the mental space is determined not by the
morphology of the neuronal tree, but by the coding system for states
of neurons.

Let us consider more advanced system of coding based on frequencies of spiking for 
neurons, e.g., Hoppensteadt,  1997.  We assign to each
neuron its frequency of spiking:
\begin{equation}
\label{SPF} \alpha= k, \; \mbox{for the frequency} \: \nu=\frac{2\pi k}{m}, \;
k=0,1,..., m-1.
\end{equation}
Such a system of coding induces the $m$-adic mental space,
$X_{\rm{mental}}={\bf Z}_m$ for any neuronal tree.
Each mental function is based on its own neuronal tree:
$T_{\rm{neuronal}}= T_{\rm{neuronal}}(f).$

\section{Model of Cognitive Psychology}

We point out that the model which is developed in this paper is a
model of neuropsychology and not at all a model of neurophysiology.
The neuronal trees under consideration are not trees for integration-propagation 
of sensory stimuli forming new mental categories at
each level of such a tree (see Khrennikov, 2002, for a general model).
 We consider neuronal trees creating
associations. As an example, let us consider a neuronal tree which
is used for representation of persons. There can be used
various hierarchical representations. We choose the "sex-representation": the
state of the root-neuron, $\star,$ gives the sex of a person:
$\alpha_0=1$ -- female, $\alpha_0=0$ -- male. Consider a branch of
this tree. Suppose that in this representation the next neuron
(after $\star)$ gives the age of a person: $\alpha_1=1$ -- young,
$\alpha_1=0$ -- not, and so on: $\alpha_2=1$ -- blond, $\alpha_2=0$
-- not, $\alpha_3=1$ -- high education/not,...

Take the ball $U_{1/2}=\{x=\alpha_0 \alpha_1...\alpha_N...:
\alpha_0=1 \}.$ This is the association of woman.

Take the ball $U_{1/4}=\{x=\alpha_0 \alpha_1...\alpha_N...:
\alpha_0=1, \alpha_1=1 \}.$ This is the association of young woman.

Take the ball $U_{1/8}=\{x=\alpha_0 \alpha_1...\alpha_N...:
\alpha_0=1, \alpha_1=1, \alpha_2=1 \}.$ This is the association of
young blond woman.

Take the ball $U_{1/16}=\{x=\alpha_0 \alpha_1...\alpha_N...:
\alpha_0=1, \alpha_1=1, \alpha_2=1, \alpha_3=1  \}.$ This is the
association of young blond woman with high education.

Take now the ball $V_{1/4}=\{x=\alpha_0 \alpha_1...\alpha_N...:
\alpha_0=0, \alpha_1=1 \}.$ This is the association of young man.

Take the union of two balls-associations: $ W= U_{1/4} \cup
V_{1/4}=\{x=\alpha_0 \alpha_1...\alpha_N...: \alpha_1=1 \}. $ This
is an idea of young person.

"Young person", $W,$ is an idea with respect to the hierarchy based
on sex. If we consider another hierarchy (based on another neuronal
tree) for that the root-neuron represents not sex, but age, then it
produces "young person" as the association: $
U_{1/2}^{\rm{age}}=\{y=\beta_0 \beta_1...\beta_N...: \beta_0=1\}. $

But $U_{1/2}^{\rm{age}}$ is not completely the same mental object as
$ W.$ The $U_{1/2}^{\rm{age}}$ is the "unisex young person" and $W$
"young person with sex."

We see that our considerations of changing of neuronal trees and
hence mental representations is similar to the choice of coordinate
systems in physics.

As was remarked at the beginning of this section, dynamics of associations and 
ideas need not be based on external stimuli (sensor or mental). Thus a neuronal tree
can be self-activated even without signals from outside, cf. with experimental results
Luczak et al., 2007.

\section{Dynamics of Associations and Ideas}

Dynamics of associations (\ref{i3}) automatically induces dynamics
of ideas
\begin{equation}
  \label{i4}
  J_{n + 1} = f(J_n).
\end{equation}
Geometrically associations are represented as bundles of branches of
the $m$-adic tree. Ideas are represented as sets of bundles. Thus
dynamics (\ref{i2}), (\ref{i3}), (\ref{i4}) are, respectively,
dynamics of branches, bundles and sets of bundles on the $m$-adic
tree.  To give examples of $f$ mapping balls onto balls, we use the
standard algebraic structure on ${\bf Z}_m$. For example, it is
known, Khrennikov 1997, 2004a, that all monomial dynamical systems
belong to this class.

We are interested in attractors of dynamical system (\ref{i4})
(these are ideas-solutions). To define attractors in the space of
ideas $X_{ID},$ we have to define a convergence in this space. We
must introduce a distance on the space of ideas (sets of
associations). Unfortunately there is a small mathematical
complication. A metric on the space of points does not induce a
metric on the space of sets that provides an adequate description of
the convergence of ideas. It is more useful to introduce a
generalization of metric, namely so called {\it pseudometric.}
\footnote{In fact, it is possible to introduce even a metric
(Hausdorff's metric) as people in general topology do. However, it
seems that this metric does not give an adequate description
 of dynamics of associations and ideas.}
Hence dynamics of ideas is a dynamics not in a metric space, but in
more general space, so called pseudometric space.

Let $(X,\rho)$ be a metric space. The distance between a point $a
\in X$ and a subset $B$ of $X$ is defined as
$$
\rho(a, B)= \inf_{b\in B} \rho(a,b)
$$
(if $B$ is a finite set, then $\rho(a, B)= \min_{b\in B}
\rho(a,b)).$

Denote by ${\rm Sub} (X)$ the system of all subsets of $X.$ {\it
Hausdorff's} distance between two sets $A$ and  $B$ belonging to
${\rm Sub} (X)$ is defined as
\begin{equation}
\label{H} \rho(A, B)= \sup_{a \in A} \rho(a, B) =\sup_{a \in A}
\inf_{b\in B} \rho(a,b).
\end{equation}
If $A$ and $B$ are finite sets, then 
$$ 
\rho(A, B)= \max_{a \in A}
\rho(a, B) =\max_{a \in A} \min_{b\in B} \rho(a,b). 
$$ 
Hausdorff's
distance $\rho$ is not a metric on the set $Y= {\rm Sub} (X).$ In
particular, $\rho(A,B)=0$ does not imply that $A=B.$ Nevertheless,
the triangle inequality $ \rho(A,B) \leq \rho(A,C) + \rho(C,B), \;
\; A,B,C \in Y,$ holds true for Hausdorff's distance.

Let $T$ be a set. A function $\rho: T \times T \to {\bf R}_+=
[0,+\infty)$ for which the triangle inequality holds true is called a
{\it pseudometric}  on $T;\; \; (T, \rho)$ is called a pseudometric
space. Hausdorff's distance is a pseudometric on the space $Y$ of
all subsets of the metric space $X;$ $(Y,\rho)$ is a pseudometric
space. The strong triangle inequality $ \rho(A,B) \leq
\max[\rho(A,C), \rho(C,B)] \; \; A,B,C \in Y, $ holds true for
Hausdorff's distance corresponding to an ultrametric $\rho$ on $X.$
In this case Hausdorff's distance $\rho$ is an {\it
ultra-pseudometric} on the set $Y=\rm{Sub}(X).$

\section{Advantages of dynamical processing of associations and ideas}

As was already mentioned, the main distinguishing feature of the dynamics of 
associations and ideas are their regularity comparing with the dynamics of 
mental states. Typically the dynamics of mental states is irregular. Numerous 
cycles and ergodic components appear and disappear depending on $m.$ 
Moreover, dynamics on more complex mental spaces (larger $m$ and 
more floors for mental trees) is more irregular than dynamics on simpler mental spaces.
Thus cognitive systems having more complex brains would have real problems with successive 
processing of information, i.e., with obtaining attractors-solutions (of course, 
under the assumption that
our model for the hierarchical dynamical processing of mental information is adequate 
to the functioning of the real brain).

Surprisingly such an irregularity for mental states induces the regular dynamics of 
associations and ideas, Khrennikov 1997, 2004a. Cycles of states disappear. They are 
hidden in balls-associations. Ergodic components are also unified into balls-associations.

Thus by using the associative representation of mental information the brain working 
as a collection of dynamical systems on hierarchical trees essentially increases the 
regularity of information processing. By our model primitive brains (having a few levels 
of mental hierarchy and rather weak networks of connections between hierarchical levels) 
are fine by working only with mental states. However,  a more complex brains {\it should form 
associations} to stabilize dynamical processing of information.

\section{Transformation of unconscious mental flows into conscious flows}
We represent a few mathematical models of the information
architecture of conscious systems $\tau,$ cf., e.g., Fodor and
Pylyshyn, 1988, Edelman, 1989, Voronkov, 2002a. We start with a quite
simple model (Model 1). This model will be developed to more complex
models which describe some essential features of human cognitive
behavior. The following sequence of cognitive models is related to
the process of evolution of the mental architecture of cognitive
systems.

 \subsection{ Model 1}

$\; \; \; \; \; $ A). The brain of $\tau$ is split into two domains:
{\it conscious} and {\it unconscious.}

B). There are two control centers, namely a {\it conscious control
center} $CC$ and an {\it unconscious control center} $UC$.

C). The main part of the unconscious domain is a {\it processing
domain} $\Pi$. Dynamical thinking processors $\pi$ are located in
$\Pi$.

\medskip

In the simplest case the outputs of some group of thinking
processors $\pi^1_{\rm{un}},\ldots,\pi_{\rm{un}}^{n}$ are always
sent to $UC$ and the outputs of another group $\pi^1_{\rm{c}}, \ldots,
\pi_{\rm{c}}^{m}$ are always sent to $CC$.\footnote{Information
produced by $\pi_{\rm{un}}$ cannot be directly used in the conscious
domain. This information circulates in the unconscious domain.
Information produced by $\pi_{\rm{c}}$ can be directly used in the
conscious domain.} The brain of $\tau$ works in the following way.

\medskip

External information is transformed by $CC$ into some
problem-idea $J_0$. The $CC$ sends $J_0$ to a thinking processor
$\pi$ located in the domain $\Pi$. Starting with $J_0$, $\pi$
produces via iteration $J_{1},\ldots, J_{N},\ldots$ an
idea-attractor $J$.

If $\pi$ = $\pi^{i}_{\rm{un}}$ (one of the unconscious-output
processors), then $J$ is transmitted to the control center $UC$.
This center sends $J$ as an initial idea $J'_0=J$ to $\Pi$ or to a
physical (unconscious) performance. In the first case some
$\pi^\prime$ (it can be conscious - output as well as unconscious -
output processor) performs iterations $J_0^\prime, J_1^\prime,
\ldots,  J_N^\prime,\ldots$ and produces a new idea-attractor
$J^\prime$. In the second case $J$ is used as a signal to some
physiological system.

If $\pi$=$\pi_{c}^{i}$ (one of conscious-output processors), then
$J$ is transmitted to the control center $CC$. This center sends $J$
as an initial idea $J_0 ^\prime = J$ to $\Pi$ (to some $\pi^\prime$)
or to a physical or mental performance (speech, writing), or to
memory. There is no additional analysis of an idea-attractor $J$
which is produced in the unconscious domain. Each attractor is
recognized by the control center $CC$ as a solution of the initial
problem $J_0,$ compared with models 2-4.

Moreover, it is natural to assume that some group of thinking
processors $\pi_{\Pi}^{1},\ldots, \pi_{\Pi}^{l}$ have their output
only inside the processing domain $\Pi$. An attractor $J$ produced
by $\pi_{\Pi}$ is transmitted neither to $CC$ nor to $UC.$ The $J$
is directly used as the initial condition by some processor $\pi$.
Finally, we obtain the  mental architecture of a brain given by Figure
2.

In this model $CC$ sends all ideas obtained from the unconscious
domain to  realization: mental or physical performance, memory
recording, transmission to $\Pi$ for a new cycle of the process of
thinking. If the intensity of the flow of information from the
unconscious domain is rather high, then such a $\tau$ can have
problems with realizations of some ideas.

\medskip

{\bf Example 9.1.} (Primitive love). Let $\tau$ be a `man' described
by this model and let $\pi = \pi_{\rm{sex}}$ be his sexual thinking
system. The image $J_0$ of a woman $\gamma$ is sent by $CC$ to
$\pi_{\rm{sex}}.$ This thinking block performs iterations $J_0, J_1,
\ldots, J_N$ and produces an idea - attractor $J$. In the simplest
case we have the pathway: $CC \rightarrow \pi_{\rm{sex}} \rightarrow
CC$ (in principle, there could be extremely complex and long
pathways, for example, $CC \rightarrow \pi_{\rm{sex}} \rightarrow
\pi^\prime \rightarrow \pi^{\prime\prime} \rightarrow UC \rightarrow
\pi^{\prime\prime\prime} \rightarrow CC$). Suppose that the idea
$J_{\rm{love}}$ = ({\small love $\gamma$}) is the attractor for
iterations starting with the image $J_0$ of a woman. Then
$J_{\rm{love}}$ is sent directly to  realization. Thus $\tau$ has no
doubts and even no craving. He performs all orders of the
unconscious domain. In fact, $CC$ can be considered as a simple
control device performing the connection with the external world.
The $\tau$ could not have mental problems. The only problem for
$\tau$ is an intensive flow of images of  women. This problem can be
solved if $\tau$ collects images and then chooses randomly an image
for realization.

\medskip

The reader may ask: Why does such a cognitive system $\tau$ need to
split mental processing into conscious and unconscious domains? The
main consequence of this splitting is that the $\tau$ does not
observe iterations of dynamical systems performing intensive
computations. The consciousness, CC, pays attention only to results
(attractors) of functioning of thinking processors. As a
consequence, the $\tau$ is not permanently disturbed by these
iterations. It can be concentrated on processing of external
information and final results of the process of thinking.

\medskip

\begin{figure}[hp]
\setlength{\unitlength}{0,8cm}
\begin{picture}(20,10)

\put(0,0){\line(1,0){14.5}} \put(2.5,0.5){\line(1,0){11}}
\put(2.5,1.5){\line(1,0){11}} \put(2.5,2.5){\line(1,0){11}}
\put(2.5,4.5){\line(1,0){11}} \put(4.5,6.5){\line(1,0){4}}
\put(4.5,10.5){\line(1,0){4}} \put(13.5,3.5){\line(1,0){1}}

\put(2.5,0.5){\line(0,1){4}} \put(13.5,0.5){\line(0,1){4}}
\put(4.5,6.5){\line(0,1){4}} \put(8.5,6.5){\line(0,1){4}}
\put(6.5,6.5){\line(0,1){4}} \put(4,2.5){\line(0,1){2}}
\put(6,2.5){\line(0,1){2}} \put(7.5,2.5){\line(0,1){2}}
\put(8.5,2.5){\line(0,1){2}} \put(9.5,2.5){\line(0,1){2}}
\put(10.5,2.5){\line(0,1){2}} \put(11.5,2.5){\line(0,1){2}}
\put(12.5,2.5){\line(0,1){2}}

\put(0,0){\vector(0,1){3.5}} \put(0,3.5){\vector(1,0){2.5}}
\put(1,2){\vector(1,0){1.5}} \put(1,2){\line(0,1){2.5}}
\put(1,5){\vector(0,-1){0.5}} \put(1,5){\vector(0,1){1}}
\put(1,6){\line(1,0){4}} \put(5,6){\vector(0,1){0.5}}

\put(3.5,4.5){\vector(0,1){0.5}} \put(6.5,4.5){\vector(0,1){0.5}}
\put(3.5,5){\line(1,0){3}} \put(5.5,6.5){\vector(0,-1){1}}
\put(6,5){\vector(0,1){1.5}} \put(3,5.5){\vector(0,-1){1}}
\put(13,5.5){\vector(0,-1){1}}

\put(3,5.5){\line(1,0){10}}

\put(7,5.5){\vector(0,-1){1}} \put(8,5.5){\vector(0,-1){1}}
\put(10,5.5){\vector(0,-1){1}}

\put(3,2){\vector(0,1){1}} \put(6.8,2){\vector(0,1){1}}
\put(7.8,3){\vector(0,-1){1}} \put(8.2,2){\vector(0,1){1}}
\put(9.8,3){\vector(0,-1){1}} \put(10.2,2){\vector(0,1){1}}
\put(10.8,2){\vector(0,1){1}} \put(13.2,2){\vector(0,1){1}}
\put(14.5,3.5){\vector(0,-1){3}} \put(14.5,0.5){\line(0,-1){0.5}}
\put(11,3.5){\vector(-1, 0){1}}

\put(3.2,4){\makebox(0,0){$\pi^{1}_{c}$}}
\put(6.8,4){\makebox(0,0){$\pi^{m}_{c}$}}
\put(8,4){\makebox(0,0){$\pi^{1}_{un}$}}
\put(10,4){\makebox(0,0){$\pi^{n}_{un}$}}
\put(11,4){\makebox(0,0){$\pi^{1}_{\Pi}$}}
\put(13,4){\makebox(0,0){$\pi^{e}_{\Pi}$}}

\put(13.5,2){\vector(1,0){4}} \put(15,10){\vector(-1,0){6.5}}
\put(4.5,9.5){\vector(-1,0){2}} \put(4.5,8.5){\vector(-1,0){3.5}}
\put(4.5,7.5){\vector(-1,0){3.5}}

\put(2,9){\makebox(0,0){Mental performance}}
\put(2,8){\makebox(0,0){Physical performance}}
\put(3,10){\makebox(0,0){Memory}}
\put(12,10.5){\makebox(0,0){External information}}
\put(16.1,2.5){\makebox(0,0){Physical}}
\put(16.1,1.5){\makebox(0,0){performance}}
\put(5.5,10){\makebox(0,0){$CC$}} \put(9,2){\makebox(0,0){$UC$}}

\put(11,2.2){\vector(0,1){0.4}} \put(12.7,2.2){\vector(0,1){0.4}}
\put(11,2.2){\line(1,0){1.7}}

\put(8,-1){\makebox(0,0){Unconscious domain}}
\end{picture}
\bigskip
\caption{Model 1 of conscious/unconscious functioning}
\end{figure}
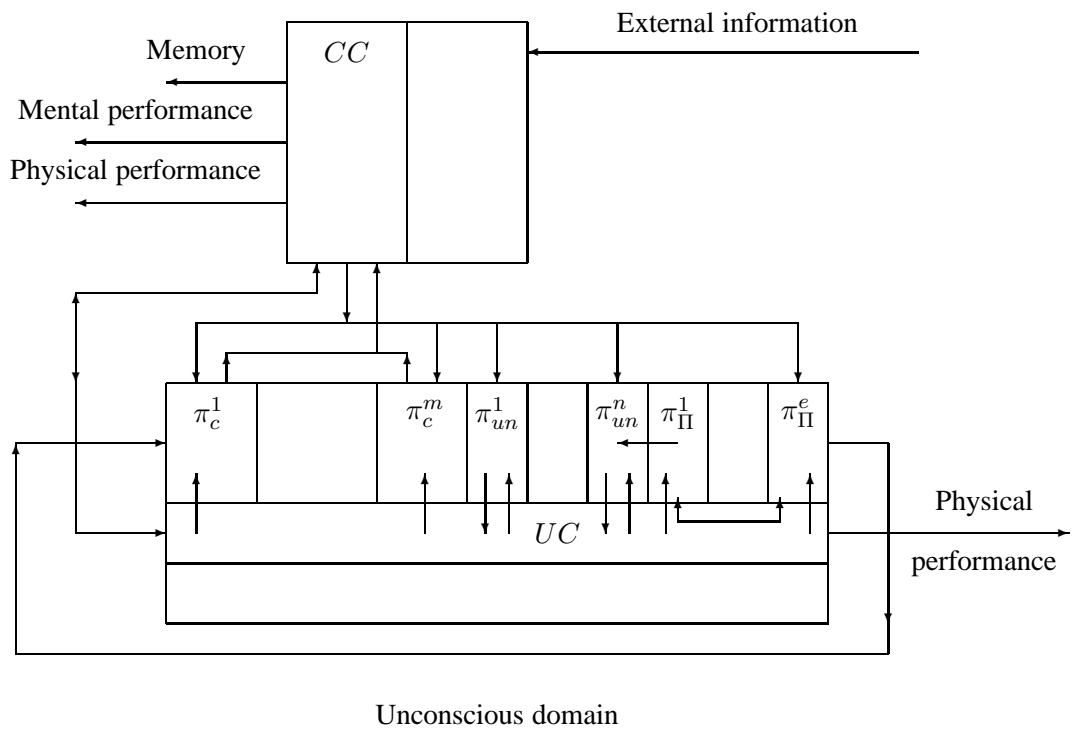

{\small  Besides the unconscious control center $UC$ and the
processing domain $\Pi$, the unconscious domain contains some other
structures (empty boxes of this picture). These additional
structures (in the conscious as well as unconscious domains) will be
introduced in more complex models. We shall also describe the
character of connections between $CC$ and $UC$. In general we need
not assume the specialization of processors $\pi$ in $\Pi:$
($\pi_{\rm{c}}^{1},\ldots, \pi_{\rm{c}}^{m})\rightarrow CC,$
($\pi_{\rm{un}}^{1},\ldots,\pi_{\rm{un}}^{n})\rightarrow UC,$
($\pi_{\Pi}^{1}, \ldots, \pi_{\Pi}^{l})\rightarrow (\pi_{\Pi}^{1},
\ldots,\pi_{\Pi}^{l}$).}

\subsection{Model 2}

One of the possibilities to improve functioning of $\tau$ is to {\it
create a queue of ideas} $J$ waiting for a realization. Thus it is
natural to assume that the conscious domain contains some {\it
collector} $Q$ in that all `waiting ideas' are gathered.

\medskip

Ideas in $Q$ must be ordered for successive  realizations. The same
order structure can be used to delete some ideas if $Q$ is complete.
Thus all conscious ideas must be classified.

\medskip

They obtain some characteristics $I(J)$ that gives a {\it measure of
interest} to an idea $J$. We may assume that  $I$ takes values in
some segment $[\delta, 1]$ (in the $m$-adic model $\delta=1/2$, see
Remark 9.1). If $I(J)=1,$ then an idea $J$ is extremely interesting
for $\tau.$ If $I(J)=\delta,$ then $\tau$ is not  at all interested
in $J.$ There exists a threshold $I_{\rm{rz}}$ of the minimal
interest for realization. If $I(J)< I_{\rm{rz}}$, then the control
center $CC$ directly deletes $J,$ despite the fact that $J$ was
produced in the unconscious domain as the solution of some problem
$J_0.$ If $I(J)\geq I_{\rm{rz}},$ then $CC$ sends the idea $J$ to
$Q.$

The $\tau$ lives in the continuously changed environment. The $\tau$
could not be concentrated on realization of only old ideas $J$ even
if they are interesting. Realizations of new ideas which are related
to the present instant of time $t$ can be more important. The
time-factor must be taken into account.

Let $l(t), \; l(0)=1,$ be some function (depending on $\tau$) which
decreases with the increasing of time $t$. Suppose that the interest
$I(t,J)$ of an idea $J$ in the queue $Q$ evolves as $$ I(t,J) = l (t
- t_0) \; I(J)\;, $$ where $I(J)$ is the value of interest of $J$ at
the instant $t_0$ of the arrival to the collector $Q$. Thus the
interest to $J$ is continuously decreasing. Finally, if $I(t,J)$
becomes less than the realization threshold $I_{\rm{rz}}$ the $J$ is
deleted from $Q$.

Quick reactions to new circumstances can be based on an
exponentially decreasing coefficient $l(t) : l(t)=e^{-Ct},$
 where a
constant $C > 0$ depends on $\tau.$\footnote{It may be that the
level of interest of $J$ evolves in a more complex way. For example,
$I(t,J)=\exp\{-C(J) t\} I(J).$ Here different ideas $J$ have
different coefficients $C(J)$ of decreasing interest.} If an idea
$J$ has an extremely high value of interest $I(J) \geq I_+$ (where
$I_+$ is a preserving threshold), then it must be realized in any
case. In our model we postulate that if $I(J) \geq I_+$, the
interest to $J$ is not changed with time: $I(t,J)=I(J)$.
\footnote{For example $I(t,J)= \exp\{ -C(J)t\} I(J)$, where $C(J)=0$
for $I(J) \geq I_+.$ So $C(J) = \alpha\; \theta (I_+ - I(J))$, where
$\alpha > 0$ is a constant (parameter of the brain) and $\theta$ is
a Heavyside function: $\theta(t)= 1, t\geq 0,$ and $\theta(t)= 0, t
< 0.$} We note that, of course, $I_+\geq I_{\rm{rz}}.$

We now describe one of the possible models for finding the value $I(J)$
of interest for an idea $J$.

The conscious domain contains a {\it database $D_i$ of ideas which
are interesting for $\tau$.} A part of this database $D_i$ was
created in the process of evolution. It is transmitted from
generation to generation (perhaps even DNA?). A part of the $D_i$ is
continuously created  on the basis of $\tau$'s experience.

The conscious domain contains a special block, {\it comparator,}
$\rm{COM}_c$ that measures the distance between two ideas, $\rho
(J_1, J_2),$ and the distance between an idea $J$ and the set $D_i$
of interesting ideas: $\rho(J,D_i)$.

At the present level of development of neurophysiology we cannot
specify a {\it mental distance} $\rho.$ Moreover, such a distance
may depend on a cognitive system or class of cognitive systems. The
hierarchic structure of the process of thinking gives some reasons
to suppose that $\rm{COM}_c$ might use the $m$-adic pseudometric
$\rho_m$ on the space of ideas $X_{ID}.$ Thus the reader may assume
that everywhere below $\rho$ is generated by the $m$-adic metric.
However, all general considerations are presented for an arbitrary
metric.

We recall that the distance between a point $b$ and a finite set $A$
is defined as $ \rho (b, A)= \min_{a\in A} \rho (b,a). $ If $J$ is
close  to some interesting idea  $L_0 \in D_i,$ then $\rho(J,D_i)$
is small. In fact, we have $ \rho(J,D_i) \leq  \rho(J,L_0), \;L_0
\in D_i. $ If $J$ is far from all interesting ideas $L \in D_i$,
then $\rho(J,D_i)$ is large. We define a measure of interest $I(J)$
as
\[
I(J)= \frac{1}{1 + \rho(J,D_i)} \;.
\]
Thus, $I(J)$ is large if  $\rho(J,D_i)$ is small; $I(J)$ is small if
$\rho(J, D_i)$ is large.

\medskip

{\bf Remark 9.1.} (The range of interest in the $m$-adic model).
Suppose that the distance $\rho$ is bounded from above:
$$
\sup_{J_1,J_2} \rho(J_1, J_2) \leq C,\; \; J_1, J_2 \in X_{ID}.
$$
Then $I(J)\geq \delta=\frac{1}{1+C}.$ In such a case `$I(J)$ is very
small' if $I(J)\approx \delta.$ The function $I(J)$ takes values in
the segment $[\delta, 1]$ (we remark that if $\rho(J,D_i)=0,$ then
$I(J)=1).$ Let $\rho$ be Hausdorff's pseudometric induced on the
space of ideas $X_{ID}$  by the $m$-adic metric  $\rho_m.$ We have
$\rho(J_1, J_2)\leq 1$ for every pair of ideas $J_1, J_2.$ Here
$\delta=1/2$ and $I(J)$ always belongs to $[1/2, 1].$ The sentence
`$I(J)$ is very small' means that $I(J)\approx 1/2$  and, as always,
`$I(J)$ is very large' means that $I(J) \approx 1.$

\medskip

There should be a connection between the level $I(J)$ of interest
and the strength of realization of $J$. Signals for mental or
physical performances of $J$ increase with increasing of $I(J)$. If,
for example, the idea  $J=\{${\small to beat this person}$\}$ then
the strength of the beat increases with increasing of $I(J)$.

In the process of memory recording the value of $I(J)$ also plays
an important role. It is natural to suppose that in working memory
the evolution of the quantity $I(t,J)$ is similar to the evolution  which was considered 
in  $Q:$ $ I(t,J)=l_{\rm{mem}} (t-t_0) I(J),$ where
$l_{\rm{mem}} (0)=1$ and $l_{\rm{mem}} (t)$ decreases with
increasing of $t$. If $I(t,J)$ becomes less than the {\it memory
preserving threshold} $I_-^{\rm{mem}}$, then $J$ is deleted from
working memory.

\medskip

{\bf Example 9.2.} (Love with interest). Let $\tau$ be a `man'
described by Model 2. In the same way as in Example 9.1 the image
$J_0$ of a woman $\gamma$ may produce the idea $J_{\rm{love}}=$
({\small love } $\gamma)$. However, $J_{\rm{love}}$ is not sent  to
realization automatically. The $\rm{COM}_c$ measures
$\rho(J_{\rm{love}}, D_i)$. Suppose that the database $D_i$ of
interesting ideas contains the idea (image) $L_{\rm{blond}}$
=({\small blond woman}). If $\gamma$ is blond, then
$\rho(J_{\rm{love}}, D_i)$ is small. So $I(J_{\rm{love}})$ is large
and $CC$ sends $J_{\rm{love}}$ to realization. However, if $\gamma$
is not blond, then $J_{\rm{love}}$ is deleted (despite the
unconscious demand $J_{\rm{love}}$). Of course, the real situation
is more complicated. Each $\tau$ has his canonical image
$L_{\rm{blond}}$. As $J_{\rm{love}} =J_{\rm{love};\gamma}$ depends
on $\gamma,$  distance $\rho(J_{\rm{love}}, D_i)$ can be essentially
different for different women $\gamma.$ Thus, for one blond woman
$\gamma$, $I(J_{\rm{love}}) \geq I_{\rm{rz}}$, but for another blond
woman $\gamma$, $I(J_{\rm{love}}) < I_{\rm{rz}}$. If there are few
blond women $\gamma_{1}, \ldots, \gamma_l$ with
$I(J_{\rm{love};\gamma_j}) \geq I_{\rm{rz}}$, then all ideas
$J_{\rm{love};\gamma_j}$ are collected in $Q$. The queue of blond
women is ordered in $Q$ due to values $I(J_{\rm{love};\gamma j})$.
If, for some $\gamma$, $I(J_{\rm{love};\gamma}) \geq I_+$, then the
level of interest toward $J_{\rm{love};\gamma}$ will not decrease with
time. The level of $I(J_{\rm{love};\gamma})$ determines the
intensity of  realization of love with $\gamma$.

\medskip

The mental architecture of `brain' in Model 2 is given by Figure 3:
{\small A new block $\rm{COM}_c$ in the conscious domain measures
the distance between an idea-attractor $J$ which has been produced
in the unconscious domain and the database $D_i$ of interesting
ideas. This distance determines the level of interest for $J:
I(J)=1/(1 + \rho (J, D_i))$. Ideas waiting for realization, $J^{1},
\ldots, J^{s}$, are collected in the special collector $Q$. They are
ordered by values of $I(J): I(J^{1}) \geq I(J^{2}) \geq \ldots \geq
I(J^s) \geq I_+$. If $I(J)\geq I_+$ (where $I_+$ is the preserving
threshold), then the value of interest of $J$ does not decrease with
time.}

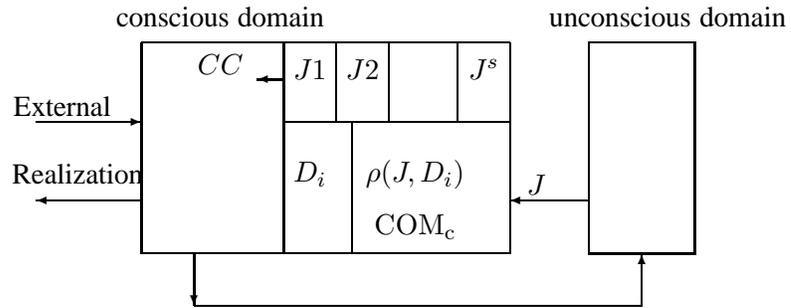
\begin{figure}[hp]
\centering \setlength{\unitlength}{0.7cm}
\begin{picture}(12,6)

\put(2,5.5){\line(1,0){7}} \put(2,5.5){\line(0,-1){4}}
\put(2,1.5){\line(1,0){7}} \put(9,5.5){\line(0,-1){4}}
\put(4.7,5.5){\line(0,-1){4}}
\put(5.7,5.5){\line(0,-1){1.5}} \put(6.7,5.5){\line(0,-1){1.5}}
\put(8,5.5){\line(0,-1){1.5}}

\put(4.7,4){\line(1,0){4.3}}

\put(6,4){\line(0,-1){2.5}}

\put(3.5,5.1){\makebox(0,0) {$CC$}} \put(5.2,5){\makebox(0,0)
{$J1$}} \put(6.2,5){\makebox(0,0) {$J2$}} \put(8.5,5){\makebox(0,0)
{$J^s$}} \put(5.2,3){\makebox(0,0) {$D_i$}}
\put(7.2,3){\makebox(0,0) {$\rho(J, D_i)$}}
\put(7.2,2){\makebox(0,0) {$\rm{COM}_c$}}

\put(10.5,1.5){\framebox(2,4)} \put(0,4){\vector(1,0){2}}
\put(2,2.5){\vector(-1,0){2}} \put(4.7,4.8){\vector(-1,0){0.5}}
\put(3,1.5){\vector(0,-1){1}} \put(3,0.5){\line(1,0){8.5}}
\put(11.5,0.5){\vector(0,1){1}} \put(10.5,2.5){\vector(-1,0){1.5}}
\put(3.5,6){\makebox(0,0) {conscious domain}}
\put(12,6){\makebox(0,0) {unconscious domain}}
\put(9.5,2.8){\makebox(0,0) {$J$}}
\put(0.5,4.3){\makebox(0,0){External}}
\put(0.8,3){\makebox(0,0){Realization}}
\end{picture}
\bigskip
\caption{Model 2 of conscious/unconscious functioning (comparative
analysis of ideas)}
\end{figure}

\subsection{Model 3}

The life of $\tau$ described by Model 2 is free of contradictions.
The $\tau$ is always oriented to realizations of the most
interesting ideas, wishes, desires. However, environment (and, in
particular, social environment) produces some constraints to
realizations of some interesting ideas.

\medskip

In a mathematical model we introduce a new quantity $F(J)$ which
describes a measure of {\it interdiction} for an idea $J$.

\medskip

 It can be
again assumed that $F(J)$ takes values in the segment $[\delta, 1].$
Ideas $J$ with small $F(J)$ have low levels of interdiction. If
$F(J)\approx \delta$, then $J$ is a `free idea'. Ideas $J$ with
large $F(J)$ have high levels of interdiction. If $F(J)\approx 1$,
then $J$ is totally forbidden.

\medskip

The interdiction function is computed in the same way as the
interest function. The conscious domain contains a {\it database
$D_f$ of forbidden ideas.} The comparator $\rm{COM}_c$ measures not
only the distance $\rho (J,D_i)$ between an idea-attractor $J$
(which has been transmitted to the conscious domain from the
unconscious domain) and the set of interesting idea $D_i,$ but also
the distance $\rho (J,D_i)$ between an idea-attractor $J$ and the
set of forbidden ideas $D_f:$
\[ \rho (J, D_f) = \min_{L\in D_f} \rho(J, L).\]
If $J$ is close to some forbidden idea $L_0$, then $\rho(J, D_f)$ is
small. If $J$ is far from all forbidden ideas $L \in D_f$, then
$\rho (J, D_f)$ is large.

We define a {\it measure of interdiction} $F(J)$ as
$$
F(J)=\frac {1}{1+\rho(J, D_f)}\;.
$$
$F(J)$ is large if $\rho(J, D_f)$ is small and $F(J)$ is small if
$\rho(J, D_f)$ is large.

\medskip

For the $m$-adic metric, $\rho(J_1, J_2)\leq 1.$ Thus $F(J)\geq
1/2.$ So $F$ takes values in the segment $[1/2,1].$ Here the
sentence `$F(J)$ is very small' means that $F(J)\approx 1/2$  and
`$F(J)$ is very large' means that $F(J) \approx 1.$

The control center $CC$ must take into account not only the level of
interest $I(J)$ of an idea $J$ but also the level of interdiction
$F(J)$ of an idea $J$. The struggle between interest $I(J)$ and
interdiction $F(J)$ induces all essential features of human
psychology. We consider a simple  model of such a struggle. For an
idea $J,$ we define {\it consistency} (between interest and
interdiction)  as $ T(J)=a I(J)-b F(J),$ where $a, b > 0$ are some
weights depending on a cognitive system $\tau.$ Some $\tau$ could
use more complex functionals for consistency. For example,
\begin{equation}
\label{T} T(J)= a I^{\alpha}(J)-b I^{\beta}(J) \; ,
\end{equation}
where $\alpha,\beta >0$, are some powers. There exists a threshold
of realization $T_{\rm{rz}}$ such that if $T(J)\geq T_{\rm{rz}}$,
then the idea $J$ is sent to the collector $Q$ for ideas waiting for
realization. If $T(J) < T_{\rm{rz}},$ then the idea $J$ is deleted.

\medskip

It is convenient to consider a special block in the conscious
domain, {\it analyzer}, $\rm{AN}_c.$

\medskip

This block contains the comparator $\rm{COM}_c$ which measures
distances $\rho(J, D_i)$ and $\rho (J,D_f)$; a computation device
which calculates measures of interest $I(J)$, interdiction $F(J)$
and consistency $T(J)$ and checks the condition $T(J) \geq T_{rz};$
a transmission device which sends $J$ to $Q$ or trash. The order in
the queue $Q$ is based on the quantity $T(J)$. It is also convenient
to introduce a block $\rm{SER}_c,$ {\it server,} in the conscious
domain which orders ideas in $Q$ with respect to values of
consistency $T(J)$.

We can again assume that there exists a threshold $T_+$ such that
ideas $J$ with $T(J)\geq T_+$ must be realized in any case. This
threshold plays the important role in the process of the time
evolution of consistency $T(t,J)$ of an idea $J$ in $Q$: $ T(t, J)=
l(t-t_0) \; T(J), $ where the coefficient $l(t)$ decreases with
increasing of $t.$ Moreover, $T(t,J)= T(J)$ if $T(J) \geq T_+$. We
note that $T_+ \geq T_{\rm {rz}}.$ It can be that interest and
interdiction evolve in different ways: $I(t,J)= l_i(t-t_0) I(J)$ and
$F(t,J)= l_f(t-t_0) F(J).$ Here $T(t,J)= a I(t,J) - b F(t,J).$ Such
a model is more realistic. A pessimist has quickly decreasing
function $l_i(t)$ and slowly decreasing function $l_f(t).$ An
optimist has slowly decreasing function $l_i(t)$ and quickly
decreasing function $l_f(t).$

{\bf Example 9.3.} (Harmonic love). Let $\tau$ be a `man' described
by Model 3. The image $J_0$ of a woman $\gamma$ is transformed by
$\pi_{\rm{sex}}$ into the idea $J_{\rm{love}, \gamma}$. Suppose that
as in Example 1.2, $D_i$ contains $L_{\rm{blond}}$ and $\gamma$ is
blond. However, $J_{\rm{love}, \gamma}$ is not sent automatically to
the queue of ideas waiting for  realization. The idea $J_{\rm{love},
\gamma}$ must be compared with the database $D_f$ of forbidden
ideas. Suppose that the idea-image $G_{\rm{tall}}=$ ({\small tall
woman}) belongs to $D_f$. If $\gamma$ is tall, then
$F(J_{\rm{love}})$ is quite large. The future of $J_{\rm{love}}$
depends on the value of the consistency functional $T(J)$ (the
relation between coefficients $a,b$ in (\ref{T}) and values $I(J)$,
$F(J))$. However, this process still does not induce doubts or
mental problems.

It seems that the consistency $T(J)$ does not determine the
intensity of realization of $J$. An extremely interesting idea is
not realized with strength that is proportional to the consistency
magnitude $T(J).$  In fact, it is realized with strength that is
proportional to the magnitude of interest $I(J).$ Moreover, larger
interdiction also implies larger strength of realization. It seems
natural to connect strength of realization with the quantity
$$
S(J) = c \; I(J)+ d \; F(J) \;,
$$
where $c, d > 0$ are some parameters of the brain.

We call $S(J)$ {\it strength} of an idea $J.$ In particular, $S(J)$
may play an important role in memory processes. We introduce a
preserving threshold $S_-^{\rm{mem}}$ (compare with the preserving
threshold $I_-^{\rm{mem}}$ in model 2). The strength $S(t,J)$ of an
idea $J$ in the working memory evolves as
$$
S(t, J)= l_{\rm{mem}} (t - t_0) S(J),
$$
 where $l_{\rm{mem}}$ is a decreasing function. If
$S(t,J) < S_-^{\rm{mem}}$, then at the instance of time $t$ the idea
$J$ is deleted from working memory.

The structure of analyzer is given by Figure 4.

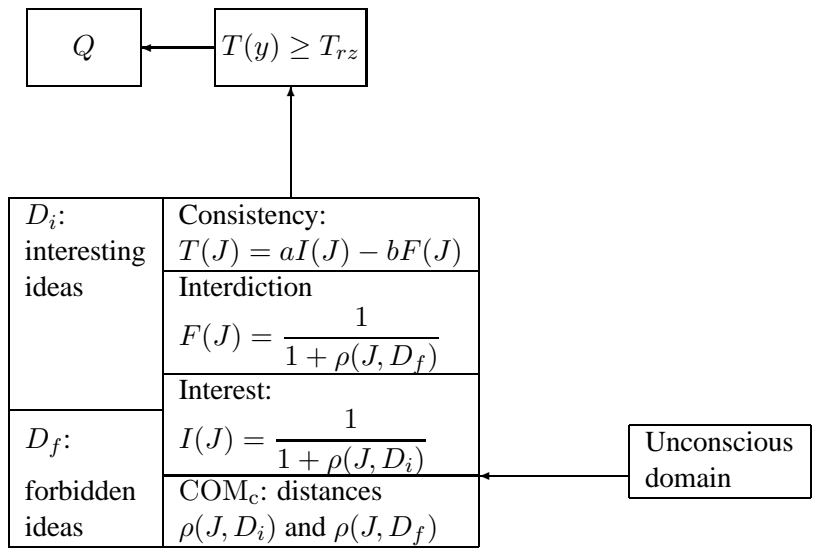
\begin{figure}[hp]
\centering \setlength{\unitlength}{1cm}

\begin{picture}(10.4,6,5)
\put(6,4){\makebox(0,0)[tr]{\begin{tabular}{|l|l|}
  \hline
  $D_i$:& Consistency:\\
  interesting & $T(J)=aI(J)-bF(J)$\\
   \cline{2-2}
   ideas & Interdiction\\
    & $F(J)=\displaystyle\frac {1}{1+\rho(J, D_f)}$\\
   \cline {2-2}
   & Interest:\\
   \cline {1-1}
   $D_f$: & $I(J)=\displaystyle\frac {1}{1+\rho (J, D_i)}$\\
   \cline {2-2}
   forbidden & $\rm{COM}_c$: distances\\
   ideas & $\rho (J, D_i)$ and $\rho (J, D_f)$\\
   \hline
\end {tabular}}}
\put(8,0.3){\vector(-1,0){2}}
\put(8,0){\makebox(0,0)[bl]{\begin{tabular}{|l|}
  \hline
  Unconscious\\
  domain\\
  \hline
  \end{tabular}}}
 \put(3.5,4){\vector(0,1){1.5}}
 \put(2.5,5.5){\framebox(2,1){$T(y)\geq T_{rz}$}}
 \put(2.5,6){\vector(-1,0){1}}
 \put(0,5.5){\framebox(1.5,1){$Q$}}
\end{picture}%

\bigskip \caption{ The structure of analyzer}
\end{figure}

{\small A cognitive system $\tau$ described by Model 3 has complex
cognitive  behavior. However, this complexity does  not imply
`mental problems'. The use of consistency functional $T(J)$ solves
the contradiction between interest and interdiction.}

\medskip

The main disadvantage of the cognitive system $\tau$ described by
Model 3 is that the analyzer $\rm{AN}_{\rm{c}}$ permits the
realization of ideas $J$ which have at the same time very high
levels of interest and interdiction (if $I(J)$ and $F(J)$ compensate
each other in the consistency function). For example, let $T(J) =
I(J)- F(J)$. If the realization threshold $T_{\rm{rz}} = 0$ the
analyzer $\rm{AN}_{\rm{c}}$ sends to the collector $Q$  totally
forbidden ideas $J$ (with $F(J) \approx 1$) having extremely high
interest $(I(J) \approx 1)$.

Such a  behavior (`a storm of cravings') can be dangerous
(especially in a group of cognitive systems with a social
structure). Therefore functioning of the analyzer $\rm{AN}_{\rm{c}}$
must be based on more complex analysis of $J$ which is not reduced
to the calculation of $T(J)$ and testing $T(J)\geq T_{\rm{rz}}.$

\subsection{Model 4}
Suppose that a cognitive system described by Model 3 improves  its
brain by introducing two new thresholds $I_{\rm{max}}$ and
$F_{\rm{max}}$. If $I(J) \geq I_{\rm{max}}$, then the idea $J$ is
extremely interesting:  $\tau$ can not simply delete $J$. If $F(J)
\geq F_{\rm{max}},$ then an idea $J$ is strongly forbidden:  $\tau$
can not simply send $J$ to $Q.$

If $J$ belongs to the {\it `domain of doubts'}
\[O_d =\{ J : I(J) \geq I_{\rm{max}} \} \bigcap \{J: F(J) \geq F_{\rm{max}} \}\]
the $\tau$ cannot take automatically (on the basis of the value of
the consistency $T(J)$) the decision on realization  of $J$.

\medskip

{\bf  Example 9.4.} (Forbidden love). Let $\tau$ be a `man'
described by Model 4. Here the image $J_0$ of a woman $\gamma$
contains not only the spatial image of $\gamma,$ but also her social
image. Suppose that the integral image $J_0$ is transformed by the
thinking block $\pi_{\rm{sex}}$ in the idea-attractor
$J_{\rm{love}}$. Suppose that, as in all previous examples, the
image $L_{\rm{blond}}$ belongs to $D_i$. Suppose that idea
$G_{\rm{soc}}$ =({\small low social level}) belongs to  $D_f.$
Suppose that both $\rho (J_{\rm{love}}, L_{\rm{blond}})$ and $\rho
(J_{\rm{love}}, G_{\rm{soc}})$ are very small. She is blond and
poor! So $I(J_{\rm{love}}) \geq I_{\rm{max}} $ (high attraction of
the woman $\gamma$ for the $\tau$) and  at the same time
$F(J_{\rm{love}})\geq F_{\rm{max}} $ (social restrictions are
important for the $\tau$). In such a situation the $\tau$ cannot
take any decision on the idea $J_{\rm{love}}$.

\section{Hidden Forbidden Wishes, Psychoanalysis}
\subsection{Hidden forbidden wishes, id\`ee fixe}
On one hand, the creation of an additional block in analyzer
$\rm{AN}_{\rm{c}}$ to perform ($I_{\rm{max}} , F_{\rm{max}} $)
analysis plays the positive role. Such a $\tau$ does not realize
automatically (via condition $T(J) \geq T_{rz}$)  dangerous ideas
$J$, despite their high attraction. On the other hand, this step in
the cognitive evolution induces hard mental problems for $\tau$. In
fact, the appearance of the domain of doubts $O_d$ in the mental
space is the origin of many psychical problems and  mental diseases.

Let $\rm{AN}_c$ find that idea $J$ belongs to $O_d$. The $\tau$ is
afraid to realize $J$ as well as to delete $J.$ The control center
$CC$ tries to perform further analysis of such a $J$. $CC$ sends
$J$ to the processing domain $\Pi$ as the initial problem for some
processor $\pi^{1}.$ If it  produces an idea-attractor $J^{1}$ which
does not belong to $O_d$, then the $\tau$ can continue normal
cognitive functioning. However, if $\pi^{1}$ produces again an idea
$J^{1}$ which belongs $O_d$, then $CC$ must continue the struggle
against this doubtful idea. In the process of such a struggle $CC$
and some processors $\pi , \pi^{1}, \pi^{2} , \ldots$ are (at least
partially) busy. An essential part of mental resources of $\tau$ is
used not for reactions to external signals, but for the struggle
with ideas $J$ belonging to $O_d$.

Typically this is a struggle with just one {\it id\`ee fixe} $J,$ see
Freud, 1933.

We can explain the origin of such an id\`ee fixe by our cognitive model.
If an idea $J$ belonging to $O_d$ has been produced by the processor
$\pi$, then it is natural that $CC$ will try again to use the same
processor $\pi$ for analyzing the idea $J$. As $ f_{\pi}(J) = J$
(the $J$ is a fixed point of the map $f_{\pi}),$ then $\pi$ starting
with $J$ will always produce the same idea $J$ (with the trivial
sequence of iterations $J, J, \ldots, J$). In general the doubtful
idea $J$ can be modified by $CC$ (for example, on the basis of new
information), $J \rightarrow J_{\rm{mod}}$. An idea $J_{\rm{mod}}$
can be considered as a perturbation of $J$: $\rho (J, J_{\rm{mod}})
< s$, where $s$ is some constant. If $s > 0$ is relatively small (so
that $J_{\rm{mod}}$ still belongs to the basis of attraction of
$J$), then iterations $J_{\rm{mod}}, J^{1}_{\rm{mod}} =
f_{\pi}(J_{\rm{mod}}), \ldots , J^{N}_{\rm{mod}} =
f_{\pi}^N(J_{\rm{mod}}), \ldots
 $
again converge to the $J$.

\medskip

How can $CC$ stop this process of the permanent work with id\`ee fixe
$J$?

\medskip

The answer to this question was given in  Freud, 1933:
investigations of roots of hysterias and some other mental
problems. By Freud id\`ee fixe $J$ is shackled by $CC$ into the
unconscious domain.

In our model, the unconscious domain contains (besides the
processing domain $\Pi$ and the unconscious control center $UC$) a
special collector $D_d$ for doubtful ideas, {\it forbidden wishes.}
After a few attempts to transform an idea $J$ belonging to the
domain of doubtful ideas $O_d$ into some non-doubtful idea, $CC$
sends $J$ to $D_d$. We remark that the domain $O_d$ is a mental
domain (a set of ideas) and $D_d$ is a `hardware domain' (a set of
chains of neurons used for saving of doubtful ideas).

What can we say about the further evolution of a doubtful idea $J$
in the collector $D_d$? It depends on a cognitive system $\tau$ (in
particular, a human individual). In the `purely normal case' the
collector $D_d$ plays just the role of a {\it churchyard for
doubtful ideas.} Such a $D_d$  has no output connections and idea
$J$ will disappear after some period of time.

\subsection{Symptoms}

However, Freud demonstrated (on the basis of hundreds of cases)
that advanced cognitive systems (such as human individuals) could
not have `purely normal behavior'. They could not perform the
complete interment of doubtful ideas in $D_d$.

In our model, the collector $D_d$ has an output connection with the
unconscious control center $UC.$ At this moment the existence of
such a connection seems to be just a disadvantage in the mental
architecture of $\tau$. It seems that such a cognitive system $\tau$
was simply  not able to develop a physical system for
100\%-isolation of the collector $D_d$. However, later we shall
demonstrate that the pathway
\begin{equation}
\label{CO} CC \rightarrow D_d \rightarrow UC \rightarrow CC
\end{equation}
has  important cognitive functions. In fact, such a connection was
specially created in the process of evolution. But we start with the
discussion on negative consequences of (\ref{CO}). Here we follow
Freud, 1933.

In our mathematical  model of Freud's theory of unconscious mind, an
idea $J \in D_d$ is sent to $UC$. The unconscious control center
$UC$ sends $J$ to one of the thinking processors $\xi$ in $\Pi$. $\xi$
performs iterations starting with $J$ as an initial idea. $\xi$
produces an idea-attractor $\tilde{J}=\lim_{N \rightarrow \infty}
J_N, J_0=J.$ In the simplest case $\xi$ sends the idea-attractor
$\tilde {J}$ to the conscious domain. The $\rm{AN}_c$ analyzes idea
$\tilde {J}.$ If $\tilde{J} \not \in O_d,$ then $ \rm{AN}_c$ sends
$\tilde{J}$ to the collector $Q$ (of ideas waiting for realization).
\footnote {Of course, there may exist more complex pathways: $CC
\rightarrow D_d \rightarrow UC \rightarrow \xi \rightarrow \xi^{1}
\ldots \rightarrow \xi^{m} \rightarrow UC \rightarrow \lambda
\rightarrow \lambda^{1} \rightarrow  \ldots \rightarrow \lambda^{k}
\rightarrow \rm{AN}_c \rightarrow Q \rightarrow CC,$  where
$\xi,..., \xi^{m}, \lambda, ...,\lambda^{k}$ are some thinking
processors.} After some period of waiting $\tilde {J}$ is sent  to
realization. \footnote{Of course, idea $\tilde {J}$ may be simply
deleted in $Q$ if there are too many ideas in the queue and the
strength $S(\tilde {J})$ of $\tilde {J}$ is not so large.} By such a
realization $CC$ deletes $J$ from the collector $Q$. However, $CC$
does not delete the root of $\tilde {J}$, namely $J$, because $J$ is
located in the unconscious domain and $CC$ is not able to control
anything in this domain. The idea $\tilde {J}$ is nothing other than a
performance of the forbidden wish $J$. Such unconscious
transformations of forbidden wishes were studied by Freud (see
Freud, 1933, for examples).

We note that if $UC$ sends a hidden forbidden wish $J$ to the same
thinking processor $\pi$ which has already generated $J$ for $CC$,
then (by the same reasons as in our previous considerations) $CC$
will again obtain the same doubtful idea $J$. Such a continuous
reproduction of id\`ee fixe can take place. This is the root of
obstinate doubtful wishes. This can imply mental deceases, because
$CC$ could not stop the struggle with id\`ee fixe even by sending it
to  $D_d.$ However, $UC$ may send $J$ to another thinking processor
$\xi \not = \pi$. Here the idea-attractor $\tilde {J}$ (which has
been produced starting with $J$ as the initial condition) differs
from $J$. This is the real transformation of the forbidden wish. In
general a new wish $\tilde {J}$ has no direct relation to the
original wish $J$. This is nothing but a {\bf symptom} of cognitive
system $\tau$, Freud, 1933.

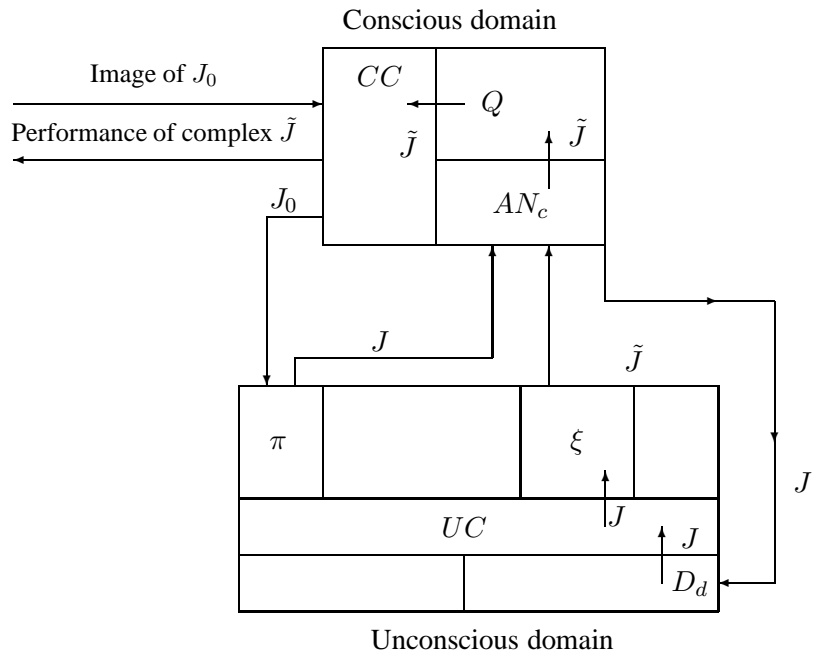
\begin{figure}[hp]
\centering \setlength{\unitlength}{0.75cm}
\begin{picture}(11,10)
\put(3.5,10){\line(1,0){2}} \put(3.5,10){\line(0,-1){3.5}}
\put(5.5,10){\line(0,-1){3.5}} \put(5.5,10){\line(1,0){3}}
\put(8.5,10){\line(0,-1){3.5}} \put(8.5,6.5){\line(-1,0){5}}
\put(2,4){\line(1,0){1.5}} \put(2,4){\line(0,-1){4}}
\put(3.5,4){\line(0,-1){2}} \put(2,2){\line(1,0){1.5}}
\put(2.7,3){\makebox(0,0){$\pi$}} \put(7,2){\line(0,1){2}}
\put(9,2){\line(0,1){2}} \put(8,3){\makebox(0,0){$\xi$}}

\put(10.5,4){\line(0,-1){4}}

\put(3.5,4){\line(1,0){7}} \put(3.5,2){\line(1,0){7}}

\put(11.5,1){\line(0,1){1}}


\put(2,0){\line(1,0){8.5}} \put(2,1){\line(1,0){8.5}}
\put(6,1){\line(0,-1){1}} \put(6,1.5){\makebox(0,0){$UC$}}
\put(10,0.5){\makebox(0,0){$D_d$}}

\put(5.5,8){\line(1,0){3}} \put(-2,9){\vector(1,0){5,5}}
\put(3.5,8){\vector(-1,0){5,5}} \put(6,9){\vector(-1,0){1}}
\put(7.5,7.5){\vector(0,1){1}} \put(3.5,7){\line(-1,0){1}}

\put(2.5,7){\vector(0,-1){3}} \put(3,4){\line(0,1){0,5}}
\put(3,4.5){\line(1,0){3.5}} \put(6.5,4.5){\vector(0,1){2}}
\put(7.5,4){\vector(0,1){2.5}}

\put(8.5,6.5){\line(0,-1){1}} \put(8.5,5.5){\vector(1,0){2}}
\put(10.5,5.5){\line(1,0){1}} \put(11.5,5.5){\vector(0,-1){2,5}}
\put(11.5,3){\line(0,-1){2,5}} \put(11.5,0.5){\vector(-1,0){1}}
\put(9.5,0.5){\vector(0,1){1}} \put(8.5,1.5){\vector(0,1){1}}

\put(6.5,9){\makebox(0,0){$Q$}} \put(7,7.2){\makebox(0,0){$AN_c$}}
\put(8,8.5){\makebox(0,0){$\tilde{J}$}}
\put(-1.5,9){\makebox(4,1){\small{Image of $J_0$}}}
\put(-1.5,8){\makebox(4,1){\small{Performance of complex
$\tilde{J}$}}} \put(3.5,10){\makebox(4.5,1){Conscious domain}}
\put(3,-1){\makebox(7,1){Unconscious domain}}
\put(2.8,7.3){\makebox(0,0){$J_0$}}
\put(4.5,4.8){\makebox(0,0){$J$}}
\put(5,8.3){\makebox(0,0){$\tilde{J}$}}
\put(4.5,9.5){\makebox(0,0){$CC$}}
\put(9,4.5){\makebox(0,0){$\tilde{J}$}}
\put(8.7,1.7){\makebox(0,0){$J$}} \put(10,1.3){\makebox(0,0){$J$}}
\put(12,2.3){\makebox(0,0){$J$}}
\end{picture}
\bigskip
\caption{Symptom induced by a forbidden wish}
\end{figure}

{\small Starting with an initial idea $J_0$ a processor $\pi$
produces an attractor $J$; analyzer $\rm{AN}_{\rm{c}}$ computes
quantities $I(J)$, $F(J)$ (measures of interest and interdiction for
the idea $J$); $\rm{AN}_{\rm{c}}$ considers $J$ as a doubtful idea:
 both measures of
interest and interdiction are too high, $I(J) \geq I_{\rm{max}} $,
$F(J) \geq F_{\rm{max}} $; $\rm{AN}_c$ sends $J$ to the collector of
doubtful ideas $D_d$; $J$ moves from $D_d$ to $UC$; $UC$ sends it to
some processor $\xi$; $\xi$ produces an attractor $\tilde{J}$.
Analyzer $\rm{AN}_c$ can recognize $\tilde {J}$ as an idea which
could be realized (depending on the distances $\rho(\tilde{J}, D_i)$
and $\rho(\tilde{J}, D_f))$ and send $\tilde {J}$ via the collector
$Q$ to realization.
 This $\tilde {J}$ is a {\it symptom} induced by $J$ (in fact, by
the initial idea $J_0)$.}

\subsection{Hysteric reactions}
In general a doubtful idea $J\in D_d$ is not only transferred into
some symptom $\tilde{J}$, but it may essentially disturb functioning
of the brain. Some thinking blocks $\pi_{\Pi}$ are directly
connected to other thinking blocks. Suppose that, for example, the
following pathway is realized: $J\in D_d\rightarrow UC \rightarrow
\pi_{\Pi} \rightarrow \pi_{\rm{c}} \rightarrow CC$. Suppose also
that ideas $\lambda$ produced by $\pi_\Pi$ play the role of
parameters for the block $\pi_{\rm{c}}: \; x_{n+1}=
f_{\rm{\pi_c}}(x_{n}, \lambda)$. Let $CC$ obtain an image $J_0$ and
send it to $\pi_{\rm{c}}$. However, instead of the normal value of
the parameter $\lambda$, the $\pi_\Pi$ sends to $\pi_{\rm{c}}$ some
abnormal value $\lambda_{\rm{ab}}$ induced by the hidden forbidden
wish $J$. The $\pi_{\rm{c}}$ produces an attractor $L_{\rm{ab}}$
which may strongly differ from the attractor $L_{\rm{norm}}$
corresponding to $\lambda_{\rm{norm}},$ the value of the parameter
produced by the processor $\pi_\Pi$ for the processor $\pi_c$  in
the absence of the hidden forbidden wish $J.$

In such a way we explain, for example, {\bf {hysteric reactions.}} A
rather innocent initial stimulus $J_0$ can induce via interference
with a doubtful idea $J \in D_d$ inadequate performance
$L_{\rm{ab}}$. We can also explain why hidden forbidden wishes may
induce {\bf {physical diseases.}} Attempting to transform $J\in D_d$
into an idea which does not induce doubts and reflections, $UC$ can
send $J$ to some thinking processor $\pi_{\rm{phys}}$ that is
responsible for some physical activity of $\tau$. We note that $UC$
considers $J$ as just a collection of mental states. This collection
of mental states has different interpretations in different thinking
systems. In particular, $J$ can correspond in $\pi_{\rm{phys}}$ to
some `bad initial condition'. The corresponding attractor
$L_{\rm{phys}}$ can paralyze the physical function ruled by
$\pi_{\rm{phys}}$.

\subsection{Feedback control based on doubtful ideas}
A cognitive system $\tau$ wants to prevent a new appearance of
forbidden wishes $J$ (collected in  $D_d$) in the conscious domain.
The brain of $\tau$ has an additional analyzer $\rm{AN}_{\rm{d}}$
(located in the unconscious domain) that must analyze nearness of an
idea attractor $L$ produced by some processor $\pi_c$  and ideas $J$
belonging to the collector of doubtful ideas $D_d.$

In our model, it is supposed that each hidden forbidden wish
$J_{\rm{fd}}$ in the collector $D_d$ still remembers a thinking
block $\pi$ which has produced $J_{\rm{fd}}.$ This simply means that
each idea $J_{\rm{fd}}$ in the $D_d$ has the label $\pi.$ Thus
$J_{\rm{fd}} =J_{\rm{fd}}(\pi)\in D_d$ is not just a collection of
mental states. There is information that these states  are related
to the dynamical system $\pi.$ The set of doubtful ideas $O$ which
are collected in the collector $D_d$ can be split into subsets
$O(\pi)$ of forbidden wishes corresponding to different thinking
systems $\pi.$ $\rm{AN}_{\rm{d}}$ contains a comparator $\rm{COM}_d$
that measures the distance between an idea-attractor $J$ which has
been produced by a thinking block $\pi$ and the set $O(\pi)$ :
$\rho(J, O(\pi)) =\min_{J_{\rm{fd}} \in O(\pi)}\rho(J,
J_{\rm{fd}})\;.$ Then $\rm{AN}_{\rm{d}}$ calculates the measure of
interdiction
\[F_d(J)=\frac{1}{1+ \rho(J,O(\pi))}. \]
If $F_d(J)$ is large $(\approx 1)$, then an idea $J$ is too close to
one of former hidden forbidden $\pi$-wishes. This idea should not be
transmitted to the conscious domain.

Each individual $\tau$ has its own {\it blocking threshold}
$F_{\rm{bl}}$: if $F_d(J) < F_{\rm{bl}}$, then $J$ is transmitted;
if $F_d(J) \geq F_{\rm{bl}}$, then $J$ is deleted. In the latter
case $J$ will never come to the conscious domain.\footnote{Analysis
in the conscious domain could demonstrate that $T(J)\geq
T_{\rm{rz}}$ and $ I(J)<I_{\rm{max}},
 F(J)<F_{\rm{max}}.$ In the absence of hidden forbidden wishes $J$ would  be  realized.}
 This threshold $F_{\rm{bl}}$ determines the degree
of blocking of the thinking processor $\pi$ by forbidden wishes. For
some individuals (having rather small values of $F_{\rm{bl}}$), a
forbidden wish $J_{\rm{fd}}$ belonging to the set $O(\pi)$ may stop
the flow of information from $\pi$ to the conscious domain. The same
$J_{\rm{fd}}$ may  play a negligible role for individuals having
rather large magnitude of $F_{\rm{bl}}$. Therefore the blocking
threshold $F_{\rm{bl}}$  is one of the important characteristics to
distinguish normal and abnormal  behaviors. We note that
$F_{\rm{bl}}$ depends on a thinking block $\pi$:
$F_{\rm{bl}}=F_{\rm{bl}}(\pi).$ Thus the same individual $\tau$ can
have the normal threshold for one thinking block $\pi,$ relatively
large $F_{\rm{bl}}(\pi)$, and abnormal degree of blocking for
another thinking block $\pi^\prime,$ relatively small
$F_{\rm{bl}}(\pi^\prime).$

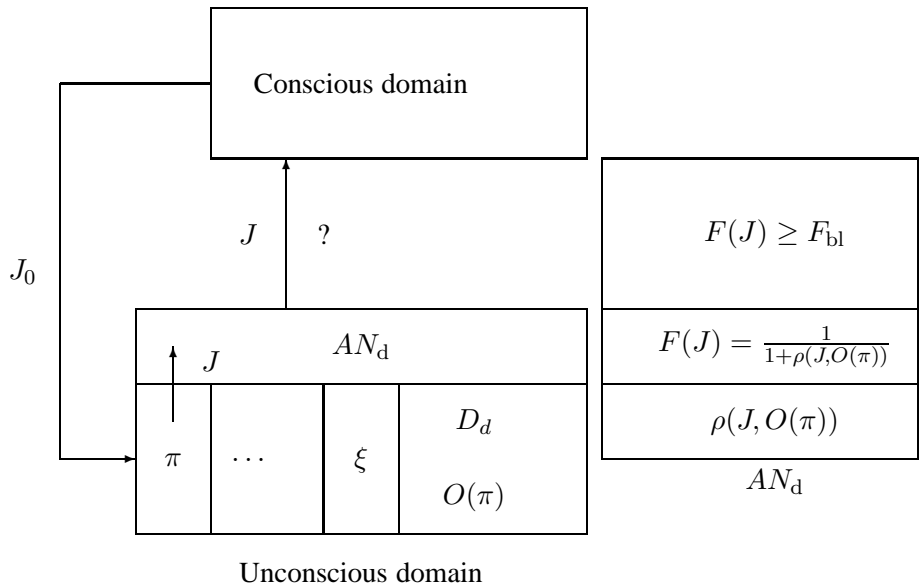
\begin{figure}[hp]
\centering \setlength{\unitlength}{1cm}

\begin{picture}(11,7)

\put(1,0){\line(1,0){6}} \put(1,2){\line(1,0){6}}
\put(1,3){\line(1,0){6}} \put(2,5){\line(1,0){5}}
\put(2,7){\line(1,0){5}} \put(7.2,1){\line(1,0){4.2}}
\put(7.2,2){\line(1,0){4.2}} \put(7.2,3){\line(1,0){4.2}}
\put(7.2,5){\line(1,0){4.2}}

\put(1,0){\line(0,1){3}} \put(7,0){\line(0,1){3}}
\put(2,0){\line(0,1){2}} \put(3.5,0){\line(0,1){2}}
\put(4.5,0){\line(0,1){2}}

\put(7.2,1){\line(0,1){4}} \put(11.4,1){\line(0,1){4}}

\put(2,5){\line(0,1){2}} \put(7,5){\line(0,1){2}}

\put(1.5,1){\makebox(0,0){$\pi$}}
\put(2.5,1){\makebox(0,0){$\ldots$}} \put(4,1){\makebox(0,0){$\xi$}}
\put(5.5,1.5){\makebox(0,0){$D_d$}}
\put(5.5,0.5){\makebox(0,0){$O(\pi)$}}

\put(4,6){\makebox(0,0){Conscious domain}}

\put(2,2.3){\makebox(0,0){$J$}} \put(2.5,4){\makebox(0,0){$J$}}
\put(3.5,4){\makebox(0,0){?}}

\put(-0.5,3.5){\makebox(0,0){$J_0$}}

\put(1.5,1.5){\vector(0,1){1}} \put(3,3){\vector(0,1){2}}

\put(2,6){\line(-1,0){2}} \put(0,6){\line(0,-1){5}}
\put(0,1){\vector(1,0){1}}

\put(9.5,1.5){\makebox(0,0){$\rho(J,O(\pi))$}}
\put(9.5,2.5){\makebox(0,0){$F(J)=\frac{1}{1+\rho(J,O(\pi))}$}}
\put(9.5,4){\makebox(0,0){$F(J)\geq F_{\rm{bl}}$}}

\put(4,-0.5){\makebox(0,0){Unconscious domain}}
\put(9.5,0.7){\makebox(0,0){$AN_{\rm{d}}$}}
\put(4,2.5){\makebox(0,0){$AN_{\rm{d}}$}}
\end{picture}
\bigskip \caption{Interference of an idea-attractor with the domain
 of hidden forbidden wishes. Internal structure of the analyzer
$\rm{AN}_{\rm{d}}.$}
\end{figure}

{\small Analyzer $\rm{AN}_{\rm{d}}$ computes the distance between
the idea-attractor $J$ (produced by a thinking block $\pi$) and the
domain $O(\pi)$  of hidden forbidden $\pi$-wishes. If this distance
is relatively small, i.e., the measure of interdiction $F_d(J)$ is
relatively large, then $J$ does not go to the conscious domain.}

\section{Consequences for neurophysiology,  neuroinformatics and cognitive sciences}

\subsection{Hierarchical models of cognition}

Such models (especially for the visual system) were discussed in a number of works:  Hubel and Wiesel, 1962, 
Bechtel and Abrahamsen, 1991, Ivanitsky, 1999, Watt  and Phillips, 2000,
Stringer and Rolls,  2002, Khrennikov, 1997, 1998a,b, 1999a,b, 2000a, b; Albeverio et
al., 1999; Dubischar et al, 1999, Voronkov, 2002a, b, Sergin,  2007.

However, the hierarchical approach has not yet become commonly accepted in 
neurophysiology,  neuroinformatics and cognitive sciences. Our mathematical model is fundamentally
hierarchical. We were able to encode hierarchy (both neuronal and mental) into  the ultrametric
geometry. Creation of a simple model of hierarchical mental space (which has natural 
coupling with  the neuronal structure of the brain) provides a mathematical basis of the 
hierarchical approach  to brain's functioning. 

On the one hand, our model needs further justification from experimental neurophysiology.
Unfortunately, at the moment there is no general consensus on the presence of hierarchical neuronal trees
in the brain. The experimental research is characterised  by diversity of opinions. Mathematical self-consistency 
of our hierarchical model might become a good stimulus for further research in neurophysiology,  neuroinformatics 
and cognitive sciences to study the hierarchy of the brain, its functioning, cognition.

On the other hand, there was created a number of theoretical hierarchical models for  processing 
of information in the brain, . However, these models were presented not on the level of mathematical 
modeling. It might be possible to use our mathematical formalism to present 
the mentioned approaches on the mathematical level.

\subsection{The problem of localization of mental functions}

This problem has been the source of permanent discussions 
in neurophysiology,  neuroinformatics and cognitive sciences
for more than one hundred years, see, e.g., Damasio, 2005,  for discussions and references.  
Our mathematical model
combines peacefully the views of both parties: the adherents of the localization  hypothesis as well 
 as the adherents of the non-localization  hypothesis. 

On the one hand, in our model each mental function $f$ is distributed over a neuronal tree. 
Branches of this tree go through different domains of the brain. Moreover, branches can contain 
even neurons belonging the spinal cord. Thus a branch can go from the cerebellum to 
the pons and then to the medulla oblongata and finally through the 
spinal cord to the conus medullaris. It is natural to include 
not only neurons, but also sensory receptors into neuronal trees 
(as belonging to the lowest level of hierarchical neuronal trees). In such a model 
delocalization of a mental function increases essentially.

Thus in our model not only the brain, but even the body participates in the thinking process.

On the other hand, the hierarchical structure of our model provides a rather sharp localization of 
a mental function $f$ (as a consequence of associative processing). 
The state $\alpha_0$ of the root-neuron
 plays the crucial role. The states of neurons of the fist level are less important and so on.

We proposed {\it the distributed model of processing of information in the brain with hierarchical
localization.}

\subsection{Binding problem}
This problem is also well known in neurophysiology,  neuroinformatics and cognitive 
sciences, see, e.g., Revonsuo and Newman,  1999, Thiele and Stoner,  2003, Zimmer et al, 2006,
for discussions and possible solutions. In our approach the binding problem 
was solved on the morphological level. The fundamental units of information processing in the brain
are not single neurons (or even localized neuronal populations), 
but hierarchical neuronal trajectories. The presence of the hierarchical structure on these
neuronal trajectories solves some problems of binding, e.g., the problem of consistency of time scales.
By operating with associations determined by short initial segments of 
hierarchical neuronal trajectories (and not with detailed mental images determined 
by the whole hierarchical neuronal trajectories) the brain uses only the top part of a 
neuronal tree. Such a computational architecture minimizes essentially the time of processing, 
cf. applications to image recognition and compression, Benois-Pineau et. al, 2001, Khrennikov 
and Kotovich, 2002,  Khrennikov, Kotovich, and  Borzistaya,  2004.

\subsection{The problem of invariance of mental images}
How can my brain recognize the image of my lovely woman? She can be dressed in different ways, she can 
express totally different emotions (from great pleasure to terrible scandals ), and so on. 
In our model the invariance is achieved through using the association-representation. Only the 
states of neurons belonging to the top levels of a neuronal tree (which is responsible for 
for the image of my lovely woman) are important in recognition. 
Temporary differences are represented 
by lower levels, cf. with $m$-adic image recognition algorithms Benois-Pineau et. al, 2001, Khrennikov 
and Kotovich, 2002,  Khrennikov, Kotovich, and  Borzistaya,  2004.

\section{Consequences for psychology and neuropsychology}
Our model gives the possibility to perform mathematical simulation of psychological behaviour.
We performed  geometrization of psychology, {\it geometro-psychology.}
By introducing a mathematical model of mental space we incorporated psychology into the same rigorous 
mathematical framework as it was done in physics by Newton and Hamilton. The crucial point is that 
geometries of physical and mental spaces differ very much. The presence of the rigid hierarchical
structure  plays the fundamental role in the $m$-adic mathematical model of mental space.

Hierarchical representations are well accepted in psychology. Our model provides 
the corresponding mathematical basis.

By coupling the $m$-adic mental space with neuronal trees we constructed a 
bridge between neurophysiology  and psychology. Thus our model can be considered 
as a contribution to neuropsychology. 

We applied the  $m$-adic mental model to mathematical modeling of Freud's psychoanalysis. 
We aware about diversity of views on Freud's psychoanalysis in modern psychology, see, e.g., 
Macmillan, 1997,  Gay, 1988,  Young-Bruehl, 1998 as well as
Stein et al.,  2006, Solms, M., 2006a, 2006b,  for debates.  
Our model supports the view which was presented in the journal 
``Neuro-psychoanalysis:'' 

It would be possible to create an ongoing dialogue with 
the aim of reconciling psychoanalytic and neuroscientific perspectives on the mind. 
This goal is based on the assumption that these two historically divided disciplines 
are ultimately pursuing the same task, namely, 
{\it ``attempt[ing] to make the complications of mental functioning intelligible 
by dissecting the function and assigning its different constituents to different 
component parts of the [mental] apparatus,''} Freud, 1900, p. 536. Notwithstanding the fact 
that psychoanalysis and neuroscience have approached this important scientific task 
from radically different perspectives, the underlying unity of purpose has become 
increasingly evident in recent years as neuroscientists have begun to investigate those 
``complications of mental functioning'' that were traditionally the preserve of psychoanalysts. 
This has produced an explosion of new insights into problems of vital interest to psychoanalysis,
but these insights have not been reconciled with existing psychoanalytic theories and models.

We can complete this manifest by the remark that neurophysiology has an essentially
higher level of the mathematical representation than traditional psychoanalysis. Therefore coupling
of psychoanalysis with neurophysiology provides new perspectives in mathematization of 
psychoanalysis. Our $m$-adic model serves precisely to such a purpose. Starting with 
a model of the neuronal structure of the brain, hierarchical neuronal trees, we created the $m$-adic
model of mental space. This model was then applied to mathematical modeling of psychoanalysis.
The $m$-adic distance on mental space is the basis of forming of measures of interest and interdiction 
and consequently hidden forbidden wishes and, finally, symptoms and hysteries. And this 
$m$-adic distance on mental space is induced by the neuronal structures -- hierarchical 
neuronal trees. There would be no psychical problems without mental hierarchy in the brain.
Psychical problems is the price for advantages of hierarchical processing of information 
in the brain.

\section{Possible consequences for medicine/psychiatry}
Our model provides an interesting explanation of differences in psychical
consequences of the same events and mental experiences. Already Freud pointed our to
the role of such differences in forming of symptoms. The same mental experience 
could play a minor role and it would be immediately forgotten by one person and it 
could be the starting point a hard psychical illness for another person, Freud, 1933.

In our model this differences in psychical reactions  to the same mental experience
are explained by the scheme of mental architecture which is based on {\it blocking 
thresholds} (such a threshold model is of course based on the presence of the ultrametric 
structure on the mental space). If this hypothesis 
were confirmed by clinical investigations, there will be opened
new ways for mental treatment. There can be developed both chemical and psychoanalytic
methods for changing the magnitudes of blocking thresholds. To develop chemical methods 
of treatment, we should find the neurophysiological basis
of blocking thresholds. There can be also developed psychoanalytic methods for 
changing the magnitudes of blocking thresholds. By special training patients can learn
to operate with blocking thresholds of lower or higher magnitudes. If a patient were able to 
make his blocking thresholds smaller, some hidden forbidden wishes would come to the consciousness.
On the one hand, his conscious mental life would become essentially more complicated. 
On the other hand, some 
symptoms would disappear. If a patient were able to 
make his blocking thresholds larger, his mental (and in particular, emotional)
behavior would become more plane. It could be important for treatment of patients with 
aggressive and destructive behavior. Such learning procedures could be based on the brain-computer
interface approach.

\section{Possible consequences for artificial life}

Investigations on artificial intelligence are oriented mainly  to creation of 
artificial systems  for motion in physical space and performing various tasks 
in this space, e.g., creation of robots.\footnote{Of course, artificial intelligence
activity is not restricted to robots. We can mention creation of chess playing machines.}
Our $m$-adic hierarchical model provides possibility for simulation of human psychology. 
In principle,
on the basis of our model artificial intellectual systems  can be created. They would live
rich emotional life: numerous interesting ideas, constraints, forbidden ideas, 
hidden forbidden wishes, feedback control based on them, symptoms and finally 
psychical problems, including hysteries. Such artificial intellectual systems
would be able to love (of course, only at the mental level), they could have various 
psychical illnesses. How can we use such {\it psychological robots?}

We can test on such psychological robots different models of mental architecture, for example, 
our hypothesis on blocking thresholds as well as our general hierarchical model.
Creation of populations of psychological robots gives the possibility for simulation of complex 
socio-psychological life.

\medskip

{\bf Conclusion.} {\it A mathematical model for hierarchical 
encoding of mental  information was created. Mental space (a mental analog
of physical space) is realized as an $m$-adic tree. Processing of mental information
is realized by dynamical systems on such a tree. Interplay between unconscious and 
conscious information flows generates interesting psychological behavior.
Consequences for neurophysiology,  neuroinformatics,
and cognitive sciences as well as for psychology and neuropsychology, and even 
medicine/psychiatry  and artificial (``psychological robots'')
were discussed. The $m$-adic hierarchical model of processing of mental information plays the role of 
the unifying mathematical basis for a number of various 
neurophysiological,  neuroinformatical and cognitive models of brain's functioning.}

\bigskip

{\bf REFERENCES}

Albeverio, S.,  Khrennikov, A. Yu.,  Kloeden, P., 1999. Memory
retrieval as a $p$-adic dynamical system. Biosystems 49, 105-115.

Amit, D., 1989. Modeling Brain Function. Cambridge Univ. Press,
Cambridge.

Ashby, R., 1952,  Design of a brain. Chapman-Hall, London.

Baars, B. J., 1997, In the theater of consciousness. The workspace
of mind. Oxford University Press, Oxford.

Bar, M., 2005. Top-down facilitation of visual object recognition. 
In Neurobiology of attention. Eds.: L. Itti, G. Rees, and J. K. Tsotsos,
Elsevier, Amsterdam, 140-145.

Bechtel, W., Abrahamsen,  A.,  1991. Connectionism and the mind.
Basil Blackwell, Cambridge.

Bechterew, W., 1911. Die Funktionen der Nervencentra. Jena, Fischer.

Benois-Pineau, J.,  Khrennikov,  A. Yu., and Kotovich, N. V., 2001.  
Segmentation of images in $p$-adic
and Euclidean metrics. Dokl. Akad. Nauk.  381, N. 5, 604-609.
English Translation: Doklady Mathematics  64, N. 3, 450-455.

Blakemore, S. J., Decety, J., 2001. From the perception of action to
the understanding of intention. Nature Reviews Neuroscience 2 (8),
561-567.

Blomberg, C., Liljenström, H., Lindahl, B.I.B., and Arhem, P.,
(Eds), 1994.Mind and Matter: Essays from Biology, Physics and
Philosophy: An Introduction, J.theor. Biol. 171.

Bredenkamp, J. (1993). Die Verknupfung verschiedener
Invarianzhypothesen im Bereich der Gedachtnispsychologie.
Zeitschrift fur Experimentelle und Angewandte Psychologie, 40,
368-385.

Chaminade, T., Meary, D., Orliaguet, J.P., Decety, J., 2001. Is
perceptual anticipation a motor simulation? A PET study Neuroreport
12 (17), 3669-3674.

Chomsky,  N.,  1963, Formal properties of grammas. Handbook of
mathematical psychology. Luce,  R. D.; Bush, R.R.; Galanter, E.;
Eds. 2, Wiley: New York, pp. 323-418.

Churchland,  P.S.,  Sejnovski,  T., 1992, The computational brain.
MITP: Cambridge.

Clark, A., 1980. Psychological models and neural mechanisms. An
examination of reductionism in psychology. Clarendon Press, Oxford.

Conte, E., Pierri,  G.P.,  Mendolicchio,  L.,   Federici, A., Zbilut
J.P., 2006. A quantum model of consciousness interfaced with a
non-Lipschitz chaotic dynamics of neural activity. Chaos, Solitons
and Fractals.

Damasio, H. and Damasio, A. R., 1989. Lesion analysis in
neuropsychology.  New-York, Oxford Univ. Press.

Damasio, A. R., 2005. Descartes' error: emotion, reason, and the human brain.
Penguin (Non-Classics).

Dragovich, B.,  and Dragovich, A.,  2006. A p-Adic model of DNA sequence and genetic code.
Electronic preprint: q-bio.GN/0607018.

Dubischar D., Gundlach, V. M., Steinkamp, O., Khrennikov, A. Yu.,
1999,  A $p$-adic model for the process of thinking disturbed by
physiological and information noise. J. Theor. Biology 197, 451-467.

Edelman, G. M.   1989. The remembered present: a biological theory
of consciousness. New York, Basic Books.

Eliasmith, C., 1996.  The third contender: a critical examination of
the dynamicist theory of cognition.  Phil. Psychology 9(4), 441-463.

Fodor, J.A. and Pylyshyn, Z. W., 1988. Connectionism and cognitive
architecture: a critical analysis,  Cognition, 280, 3--17.

Freud, S, 1900. The interpretation of dreams. Standard Edition, 4 and 5. 

Freud, S., 1933. New introductory lectures on psychoanalysis. New
York, Norton.

Frith, C.D., Frith, U., 1999, Cognitive psychology - Interacting
minds - A biological basis. Science 286 (5445), 1692-1695. 

Fuster, J.M.D., 1997. The prefrontal cortex: anatomy, physiology,
and neuropsychology of the frontal lobe. Philadelphia,
Lippincott-Raven.

Gay, P., 1988. Freud: A life for our time. W.W. Norton, NY.

Geissler, H.-G., Klix, F., and Scheidereiter, U. (1978). Visual
recognition of serial structure: Evidence of a two-stage scanning
model. In E. L. J. Leeuwenberg and H. F. J. M. Buffart (Eds.),
Formal theories of perception (pp. 299-314). Chichester: John Wiley.

Geissler, H.-G. and Puffe, M. (1982). Item recognition and no end:
Representation format and processing strategies. In H.-G. Geissler
and Petzold (Eds.), Psychophysical judgment and the process of
perception (pp. 270-281). Amsterdam: North-Holland.

Geissler, H.-G. (1983). The Inferential Basis of Classification:
From perceptual to memory code systems. Part 1: Theory. In H.-G.
Geissler, H. F. Buffart, E. L. Leeuwenberg, and V. Sarris (Eds.),
Modern issues in perception (pp. 87-105). Amsterdam: North-Holland.

Geissler, H.-G. (1985). Zeitquantenhypothese zur Struktur
ultraschneller Gedachtnisprozesse. Zeitschrift fur Psychologie, 193,
347-362. Geissler, H.-G. (1985b). Sources of seeming redundancy in
temporally quantized information processing. In G. d'Ydewalle (Ed.),
Proceedings of the XXIII International Congress of Psychology of the
I.U.Psy.S., Volume 3 (pp. 199-228). Amsterdam: North-Holland.

Geissler, H.-G. (1987). The temporal architecture of central
information processing: Evidence for a tentative time-quantum model.
Psychological Research, 49, 99-106. Geissler, H.-G. (1990).
Foundations of quantized processing. In H.-G. Geissler (Ed.),
Psychophysical explorations of mental structures (pp. 193-210).
Gottingen, Germany: Hogrefe and Huber Publishers.

Geissler, H.-G. (1992). New magic numbers in mental activity: On a
taxonomic system for critical time periods. In H.-G. Geissler, S. W.
Link, and J. T. Townsend (Eds.): Cognition, information processing
and psychophysics (pp. 293-321). Hillsdale, NJ: Erlbaum.

Geissler, H.-G. and Kompass, R. (1999). Psychophysical time units
and the band structure of brain oscillations. 15th Annual Meeting of
the International Society for Psychophysics, 7-12.

Geissler, H.-G., Schebera, F.-U., and Kompass, R. (1999).
Ultra-precise quantal timing: Evidence from simultaneity thresholds
in long-range apparent movement. Perception and Psychophysics, 6,
707-726.

Geissler, H.-G., and Kompass, R. (2001). Temporal constraints in
binding? Evidence from quantal state transitions in perception.
Visual Cognition, 8, 679-696.

Hopfield,  J. J., 1982. Neural networks and physical systems with
emergent collective computational abilities. Proc. Natl. Acad. Sci.
USA 79,  1554-2558.

Hoppensteadt, F. C., 1997.  An introduction to the mathematics of neurons:
modeling in the frequency domain.  Cambridge Univ. Press,
New York.

Hubel, D., and Wiesel, T.,  1962. Receptive fields, binocular interaction and 
functional architecture in the cat's visula cortex. J. Physiol. 160, 106-154.                                         

Ivanitsky, A.M., 1999. Brain's physiology and the origin of the human's 
subjective world. J. of High Nerves Activity 49, N. 5, 707-713.

Khrennikov, A. Yu. , 1997.  Non-Archimedean analysis: quantum
paradoxes, dynamical systems and biological models. Kluwer,
Dordrecht.

Khrennikov, A. Yu.,  1998a.  Human subconscious as the $p$-adic
dynamical system. J. of Theor. Biology 193, 179-196.

Khrennikov, A. Yu.,  1998b.  $p$-adic dynamical systems: description
of concurrent struggle in biological population with limited growth.
Dokl. Akad. Nauk 361, 752.

Khrennikov, A. Yu., 1999a, Description of the operation of the human
subconscious by means of $p$-adic dynamical systems. Dokl. Akad.
Nauk 365, 458-460.

Khrennikov, A. Yu., 2000a,   $p$-adic discrete dynamical systems and
collective behaviour of information states in cognitive models.
Discrete Dynamics in Nature and Society 5, 59-69.

Khrennikov,  A. Yu., 2000b.  Classical and quantum mechanics on
$p$-adic trees of ideas. BioSystems   56, 95-120.

Khrennikov A.Yu., 2002. Classical and quantum mental models and
Freud's theory of unconscious mind. Series Math. Modelling in
Phys., Engineering and Cognitive sciences, 1. V\"axj\"o Univ.
Press, V\"axj\"o.

Khrennikov, A. Yu.,  and Kotovich, N.V., 2002. Representation  and compression of images
with the aid of the $m$-adic coordinate system. Dokl. Akad. Nauk. 387 N. 2,
159-163.

Khrennikov, A.Yu.,  2004a, Information dynamics in cognitive,
psychological, social,  and anomalous phenomena. Kluwer, Dordreht.

Khrennikov, A. Yu., 2004b, Probabilistic pathway representation of
cognitive information. J. Theor. Biology 231, 597-613.

Khrennikov, A. Yu.,  Kotovich, N. V., Borzistaya,  E. L.,  2004. Compression of
images with the aid of representation by $p$-adic maps and
approximation by Mahler's polynomials. Dokl. Akad. Nauk
396, N 3, 305-308. English Translation: Doklady
Mathematics 69 N 3, 373-377.

Klix, F., and van der Meer, E. (1978). Analogical reasoning - an
approach to mechanisms underlying human intelligence performances.
In F. Klix (Ed.), Human and artificial Intelligence (p. 212).
Berlin: Deutscher Verlag der Wissenschaften.

Kristofferson, M. W. (1972). Effects of practice on
character-classification performance. Canadian Journal of
Psychology, 26, 540-560.

Kristofferson, A. B. (1980). A quantal step function in duration
discrimination. Perception and Psychophysics, 27, 300-306.

Kristofferson, A. B. (1990). Timing mechanisms and the threshold for
duration. In Geissler, H.-G. (Ed., in collaboration with M. H.
Muller  and  W. Prinz), Psychophysical explorations of mental
structures (pp. 269-277). Toronto: Hogrefe  and  Huber Publishers.

Luczak, A., Bartho, P., Marguet, S. L., Buzsaki, G., Hariis, K.D,
2007, Neocortical spontaneous activity in vivo: cellular
heterogeneity and sequential structure. Preprint of CMBN, Rutgers
University.

Macmillan, M., 1997. The Completed Arc: Freud Evaluated. MIT Press., 
Cambridge MA.

Murtagh, 2004. On ultrametricity, data coding, and computation. Journal of 
Classification 21, 167-184.

Oztop, E., Wolpert, D., Kawato, M. 2005,  Mental state inference
using visual control parameters. Cognitive Brain Research 22 (2),
129-151.

Pitk\"anen, M., 1998. TGD inspired theory of consiousness with
applications to biosystems. Electronic book:
http://www.physics.helsinki.fi/~matpitka/cbookI.html

Pitk\"anen,  M. 2006. Could genetic code be
understood number theoretically? 
Electronic preprint: www.helsinki.fi/~matpitka/pdfpool/genenumber.pdf.

Revonsuo, A and Newman, J., 1999. Binding and Consciousness. 
Consciousness and Cognition 8, 123-127.

Sergin, V. Ja., 2007. Biologically reasonable model of visual perception: 
hierarchy of unifying sensor characteristics. In  Neuroinformatics-2007,
Lectures on Neuroinformatics, part 2. Ed.: Yu. V. Tumenzev, 
Moscow Institute of Engineering and Physics Press, Moscow, 77-120.

Smythies, J.R.,  1970. Brain mechanisms and behaviour. Blackwell Sc.
Publ.: Oxford.

Solms, M., 2002. An Introduction to the Neuroscientific Works of Freud. 
In The Pre-Psychoanalytic Writings of Sigmund Freud. Eds.: G.
van der Vijver and F. Geerardyn. Karnac, London, 25-26.

Solms, M., 2006a. Putting the psyche into neuropsychology. Psychologist 19 (9),
538-539.

Solms, M., 2006b. Sigmund Freud today. Psychoanalysis and neuroscience in dialogue. 
Psyche-Zeiteschrift fur Psychoanalyse und ihre Anwendungen 60 (9-10),
829-859.

Stein, D. J., Solms, M., van Honk, J.,  2006. 
The cognitive-affective neuroscience of the unconscious. 
CNS Spectrums 11 (8),  580-583.

Stringer, S.M., and Rolls, E., 2002. Invariant object recognition 
in the visual system with novel views of 3D objests. Neural. Comput.
14, 2585-2596.

Strogatz,  S. H., 1994.  Nonlinear dynamics and chaos with
applications to physics, biology, chemistry, and engineering.
Addison Wesley, Reading, Mass.

Teghtsoonian, R. (1971): On the exponents in Stevens' law and on the
constant in Ekman's law. Psychological Review, 78, 71 - 80.

Thiele, A., and Stoner, G., 2003. 
Neuronal synchrony does not correlate with motion coherence in cortical area MT. 
Nature 421, 366-370.

van Gelder, T., Port, R., 1995. It's about time: Overview of the
dynamical approach to cognition. in  Mind as motion: Explorations in
the dynamics of cognition. Ed.: T. van Gelder, R. Port. MITP,
Cambridge, Mass, 1-43.

van Gelder, T., 1995. What might cognition be, if not computation?
J. of Philosophy 91, 345-381.

Vladimirov, V. S., Volovich,  I. V., and Zelenov,  E. I., 1994,
$p$-adic Analysis and  Mathematical Physics. World Scientific Publ.,
Singapore.

Voronkov, G.S., 2002a. Information and brain: viewpoint of
neurophysiolog. Neurocomputers: development and applications N 1-2,
79-88.

Voronkov, G.S.,  2002b.  Why is the perceived visual world
non-mirror? Int. J. of Psychophysiology 34, 124-131.

Watt, R.J., and Phillips, W.A., 2000. The function of dynamical grouping in vision.
Trends Cogn. Sc. 4, 447-454.

Young-Bruehl, E.,  1998.
Subject to Biography. Harvard University Press, Boston.

Zimmer, H.,  Mecklinger, A.,  and  Lindenberger U., 2006. Handbook of binding and memory.
Perspectives from cognitive neuroscience. Oxford Univ. Press, Oxford.

\end{document}